\documentclass[aps,eqsecnum]{revtex4}
\usepackage[T1]{fontenc}
\usepackage[latin1]{inputenc}
\usepackage{graphicx}
\usepackage{amssymb}

\makeatletter

\usepackage{graphicx}
\usepackage{graphicx}
\usepackage{bm}

\makeatother
\begin{document}

\title{Critical fluctuations and entanglement in the \\
 nondegenerate parametric oscillator}

\author{K.~Dechoum$^{1,2}$, P.~D.~Drummond$^{1}$, S.~Chaturvedi$^{3}$, and M. D. Reid$^{1}$}

\affiliation{$^{1}$Australian Centre for Quantum-Atom Optics, The University of Queensland, St Lucia 4067, Queensland, Australia \\
 $^{2}$ Instituto de F\'{\i}sica da Universidade Federal Fluminense, Boa Viagem Cep.:24210-340, Niter\'{o}i-RJ, Brazil
\\
 $^{3}$ School of Physics, University of Hyderabad, Hyderabad 500046, India }

\begin{abstract}
We present a fully quantum mechanical treatment of the nondegenerate optical parametric oscillator both below and near threshold.
This is a non-equilibrium quantum system with a critical point phase-transition, that is also known to exhibit strong yet
easily observed squeezing and quantum entanglement. Our treatment makes use of the positive P-representation and goes beyond
the usual linearized theory. We compare our analytical results with numerical simulations and find excellent agreement. We
also carry out a detailed comparison of our results with those obtained from stochastic electrodynamics, a theory obtained
by truncating the equation of motion for the Wigner function, with a view to locating regions of agreement and disagreement
between the two.  We calculate commonly used measures of quantum behavior including entanglement, squeezing and EPR correlations
as well as higher order tripartite correlations, and show how these are modified as the critical point is approached. In
general, the critical fluctuations represent an ultimate limit to the possible entanglement that can be achieved in a nondegenerate
parametric oscillator. 
\end{abstract}
\maketitle

\section{Introduction}

Nonlinear optical devices such as optical parametric oscillators (OPO) and optical parametric amplifiers (OPA) \cite{Yariv}
have been studied in the last forty years for providing fundamental tests of quantum mechanics, as well as their technological
applications in areas such as frequency conversion, low noise optical measurement, and cryptography. The light beams emitted
by these devices are characterized by a large amount of squeezing \cite{WKHW86}, significant quantum intensity correlations
\cite{HHRG87}, and very short correlation times between the conjugate beams \cite{Hong}. The entangled nature of the photons
in the down converted light has been instrumental in providing experimental demonstrations\cite{Ou,yhang,silber,bow} of
the original Einstein-Podolsky-Rosen paradox and other nonclassical features of quantum mechanics. In this paper, we extend
the usual linear theory of the nondegenerate OPO to include nonlinear effects characteristic of the onset of critical fluctuations
near threshold, which is the physical feature that ultimately limits the maximum squeezing and entanglement available.

As a fundamental application of these results, we point out that in 1935, Einstein, Podolsky and Rosen~\cite{epr} (EPR)
presented their famous argument which demonstrates that local realism is inconsistent with the completeness of quantum mechanics.
Their argument concerned two spatially separated particles with perfectly correlated positions and momenta, as predicted
by quantum mechanics. Related correlations for quadrature phase operators have been studied\cite{eprquad,eprr,rd} and experimentally
confirmed for the output fields of the nondegenerate parametric oscillator, both below\cite{Ou,yhang} and above\cite{Pfister}
threshold. The regime of study of these correlations however has been so far confined to regimes of operation where the quantum
fluctuations will be small so that a linearized analysis is valid. 

Closely linked with the phenomenon of EPR correlations is that of entanglement, a key feature enabling many potential applications
in the field of quantum information. Criteria for proving entanglement using continuous variable (quadrature phase amplitude)
measurements have been developed by Simon and Duan et al\cite{content}. Recent experiments\cite{bow,cventnat,cventpol}
have measured such continuous variable entanglement but again the studies are limited to the regime of stable, linearizable
quantum fluctuations. In this regime Gaussian statistics apply, and the criterion developed can be shown\cite{content} to
be both a necessary and sufficient condition for entanglement in this case.

It is known from earlier theoretical analysis \cite{TE,KTE} of the optical properties of nonlinear crystals that, in the
linearized or Gaussian regime, a local realistic theory based on the Wigner phase space representation gives the same results
for the correlations between signal and idler light beams produced in nonlinear crystals\cite{TE,KTE}. While this is also
true of many correlations in second harmonic generation \cite{Olsen}, there are instances where significant differences
exist between the predictions of the two theories \cite{Dechoum}. Here we calculate the EPR and entanglement measures for
non-Gaussian fields, in precisely the type of environment where non-Gaussian behavior is expected to occur experimentally
- that is, by considering nonlinear corrections to the usual linearized approximations used to treat the OPO below threshold.

In two recent papers \cite{CDD} we have carried out a fully quantum mechanical analysis of nonlinear effects and critical
fluctuations in a degenerate OPO using the positive P-representation, and have investigated the squeezing spectra and triple
correlations in this system both analytically as well as numerically. In particular, we have shown that, in this case, while
the full quantum theory and the semi-classical theory disagree strongly far below threshold, there is a surprising agreement
between the two close to the threshold where quantum fluctuations are quite intense, characteristic of a mixed state of light
in this limit. 

The aim of the present work is to present a similar analysis for the case of a nondegenerate optical parametric oscillator.
Both the quantum mechanical and semi-classical analyses are carried out in parallel and are compared with exact numerical
simulations. Special attention is paid to the behavior of this system close to the critical point to ascertain the limits
of entanglement, EPR correlations and squeezing in this regime. We find that entanglement is optimised just below the critical
point for output mode entanglement and squeezing, while the optimum internal squeezing and entanglement is achieved just
above threshold.

\section{Hamiltonian and stochastic equations }

We consider here the standard model for three modes coupled by a nonlinear crystal inside a Fabry-Perot interferometer with
allowance made for coherent pumping and damping due to cavity losses. This model implies certain restrictions on mode-spacing
for its validity, since we assume only these three modes are excited.

\subsection{Hamiltonian}

The Heisenberg-picture Hamiltonian that describes this open system is given by\cite{Graham,McNeil,rd}\begin{eqnarray}
\hat{H} & = & \sum_{i=0}^{2}\hbar\omega_{i}\hat{a}_{i}^{\dagger}\hat{a}_{i}+i\hbar\chi\left(\hat{a}_{1}^{\dagger}\hat{a}_{2}^{\dagger}\hat{a}_{0}-\hat{a}_{1}\hat{a}_{2}\hat{a}_{0}^{\dagger}\right)\nonumber \\
 &  & +i\hbar\left({\mathcal{E}}e^{-i\omega_{0}t}\hat{a}_{0}^{\dagger}-{\mathcal{E}}^{*}e^{i\omega_{0}t}\hat{a}_{0}\right)\nonumber \\
 &  & +\sum_{i=0}^{2}\left(\hat{a}_{i}\hat{\Gamma}_{i}^{\dagger}+\hat{a}_{i}^{\dagger}\hat{\Gamma}_{i}\right)\label{a1}\end{eqnarray}
 Here ${\mathcal{E}}$ represents the external input field at a frequency $\omega_{0}$, with $\hat{a}_{0}$, $\hat{a}_{1}$
and $\hat{a}_{2}$ represent the pump, the signal and the idler intra-cavity modes at frequencies $\omega_{0}$, $\omega_{1}$
and $\omega_{2}$ respectively, where $\omega_{0}=\omega_{1}+\omega_{2}$. The terms $\hat{\Gamma}_{i}$ represent reservoir
operators and $\chi$ denotes the nonlinear coupling constant due to the second order polarizability of the nonlinear crystal. 

This is a driven system far from thermal equilibrium, so it is not appropriate to assume a canonical ensemble. Instead, the
density matrix must be calculated as the solution of a master equation in the Schroedinger picture. For simplicity, we transform
to a rotating frame in which the free-field time-evolution is removed. The master equation for the reduced density operator,
obtained after the elimination of the reservoirs using standard techniques\cite{14}, is given by

\begin{eqnarray}
\frac{\partial\hat{\rho}}{\partial t} & = & \chi\left[\hat{a}_{1}^{\dagger}\hat{a}_{2}^{\dagger}\hat{a}_{0}-\hat{a}_{1}\hat{a}_{2}\hat{a}_{0}^{\dagger},\hat{\rho}\right]+\mathcal{E}\left[\hat{a}_{0}^{\dagger}-\hat{a}_{0},\hat{\rho}\right]\nonumber \\
 &  & +\sum_{i=0}^{2}\gamma_{i}\left(2\hat{a}_{i}\hat{\rho}\hat{a}_{i}^{\dagger}-\hat{a}_{i}^{\dagger}\hat{a}_{i}\hat{\rho}-\hat{\rho}\hat{a}_{i}^{\dagger}\hat{a}_{i}\right)\label{a2}\end{eqnarray}

\noindent where $\gamma_{i}$ are the damping rates for the mode amplitudes. For simplicity, we assume that $\gamma_{1}=\gamma_{2}=\gamma$
throughout this paper.

To handle master equations such as this it proves convenient to transform them into c-number Fokker-Planck equations or equivalently
into stochastic equations using operator representation theory. Here, as in our earlier works, we use the positive P-representation
for this purpose, and we also compare these results with the commonly used semiclassical truncation of the Wigner representation.

\subsubsection{Classical critical point}

The classical approximation, where all fluctuations are neglected, is obtained by simply assuming that all operator mean
values factorize. This gives us the classical nonlinear-optical equations for $\alpha_{i}=\langle\hat{a}_{i}\rangle$ in
the form:\begin{eqnarray}
\frac{\partial\alpha_{0}}{\partial t} & = & \mathcal{E}-\gamma_{0}\alpha_{0}-\chi\alpha_{1}\alpha_{2}\nonumber \\
\frac{\partial\alpha_{i}}{\partial t} & = & -\gamma\alpha_{i}+\chi\alpha_{3-i}^{*}\alpha_{0}\label{eq:classical}\end{eqnarray}

For small driving fields, the stable classical below-threshold solutions are $\alpha_{0}=\mathcal{E}/\gamma_{0}$ and $\alpha_{1}=\alpha_{2}=0$
. There is a classical threshold or critical point at $\mathcal{E}=\mathcal{E}_{c}=\gamma\gamma_{0}/\chi$. Above this threshold,
the driving field is clamped at $\alpha_{0}=\mathcal{E}_{c}/\gamma_{0}$, while the signal and idler intensities increase
linearly with the input field $\mathcal{E}$. This paper deals with the near threshold and sub-threshold regime.

\subsection{The positive P-representation}

Using the positive P-representation\cite{+P}, we can include correlations and fluctuations by expanding the density matrix
describing the system in an off-diagonal coherent state basis as

\begin{equation}
\hat{\rho}=\int_{\mathcal{{D}}}\frac{|\bm\alpha\rangle\langle\left(\bm\alpha^{+}\right)^{*}|}{\langle\left(\bm\alpha^{+}\right)^{*}|\bm\alpha\rangle}P_{+}(\bm\alpha,\bm\alpha^{+})d^{6}\bm\alpha d^{6}\bm\alpha^{+}\label{a3}\end{equation}

\noindent where $\bm\alpha\equiv\left(\alpha_{0},\alpha_{1},\alpha_{2}\right)$ and $\bm\alpha^{+}\equiv\left(\alpha_{0}^{+},\alpha_{1}^{+},\alpha_{2}^{+}\right)$
are two independent triplets of complex variables. The function $P(\bm\alpha,\bm\alpha^{+})$ can be understood as a positive
phase space distribution and, by virtue of $(\ref{a3})$, satisfies the following Fokker-Planck equation\cite{rd} (assuming
that boundary terms vanish on partial integration):\begin{eqnarray}
\frac{\partial P_{+}}{\partial t} & = & \left\{ \frac{\partial}{\partial\alpha_{0}}\left[\gamma_{0}\alpha_{0}+\chi\alpha_{1}\alpha_{2}-{\mathcal{E}}\right]\right.\nonumber \\
 &  & +\frac{\partial}{\partial\alpha_{0}^{+}}\left[\gamma_{0}\alpha_{0}^{+}+\chi\alpha_{1}^{+}\alpha_{2}^{+}-{\mathcal{E}}\right]\nonumber \\
 &  & +\frac{\partial}{\partial\alpha_{1}}\left[\gamma_{1}\alpha_{1}-\chi\alpha_{0}\alpha_{2}^{+}\right]+\frac{\partial}{\partial\alpha_{1}^{+}}\left[\gamma_{1}\alpha_{1}^{+}-\chi\alpha_{0}^{+}\alpha_{2}\right]\nonumber \\
 &  & +\frac{\partial}{\partial\alpha_{2}}\left[\gamma_{2}\alpha_{2}-\chi\alpha_{0}\alpha_{1}^{+}\right]+\frac{\partial}{\partial\alpha_{2}^{+}}\left[\gamma_{2}\alpha_{2}^{+}-\chi\alpha_{0}^{+}\alpha_{1}\right]\nonumber \\
 &  & \left.+\frac{\partial^{2}}{\partial\alpha_{1}\partial\alpha_{2}}(\chi\alpha_{0})+\frac{\partial^{2}}{\partial\alpha_{1}^{+}\partial\alpha_{2}^{+}}(\chi\alpha_{0}^{+})\right\} P_{+}(\alpha,\alpha^{+},t)\,\,.\nonumber \\
\end{eqnarray}

We note here that boundary terms are found to be exponentially suppressed for large damping\cite{GGD-Validity} - ie, $\gamma_{i}\gg\chi$,
which corresponds to typical experimental conditions for realistic OPO's in current use.

This can equivalently be written as the following set of It\^{o} stochastic equations \cite{Arnold}\begin{eqnarray}
d\alpha_{0} & = & \left({\mathcal{E}}-\gamma_{0}\alpha_{0}-\chi\alpha_{1}\alpha_{2}\right)dt\nonumber \\
d\alpha_{0}^{+} & = & \left({\mathcal{E}}^{*}-\gamma_{0}\alpha_{0}^{+}-\chi\alpha_{1}^{+}\alpha_{2}^{+}\right)dt\nonumber \\
d\alpha_{1} & = & \left(-\gamma_{1}\alpha_{1}+\chi\alpha_{2}^{+}\alpha_{0}\right)dt+\left(\chi\alpha_{0}\right)^{1/2}dW_{1}\nonumber \\
d\alpha_{1}^{+} & = & \left(-\gamma_{1}\alpha_{1}^{+}+\chi\alpha_{2}\alpha_{0}^{+}\right)dt+\left(\chi\alpha_{0}^{+}\right)^{1/2}dW_{1}^{+}\nonumber \\
d\alpha_{2} & = & \left(-\gamma_{2}\alpha_{2}+\chi\alpha_{1}^{+}\alpha_{0}\right)dt+\left(\chi\alpha_{0}\right)^{1/2}dW_{2}\nonumber \\
d\alpha_{2}^{+} & = & \left(-\gamma_{2}\alpha_{2}^{+}+\chi\alpha_{1}\alpha_{0}^{+}\right)dt+\left(\chi\alpha_{0}^{+}\right)^{1/2}dW_{2}^{+}\nonumber \\
\label{a6}\end{eqnarray}

where \begin{eqnarray}
\langle dW_{1}\rangle & = & \langle dW_{2}\rangle=0\nonumber \\
\langle dW_{1}dW_{2}\rangle & = & \langle dW_{1}^{+}dW_{2}^{+}\rangle=dt\label{a7}\end{eqnarray}
 with all other noise correlations vanishing. These equations imply that $\langle\alpha_{i}\alpha_{i}^{\dagger}\rangle=\langle\widehat{n}_{i}\rangle=0$
when there is no driving field, as physically expected for a vacuum state in a normally-ordered representation.

Numerical simulations of these stochastic trajectories confirms the assumption of asymptotically vanishing boundary terms
for the parameters we use, as the trajectories are strongly bounded to a compact domain. At smaller damping rates, it would
become important to include stochastic gauge terms\cite{GaugeP} in the equations to eliminate boundary terms, but this was
not found to be necessary in these calculations. In other words, while boundary terms are potentially present, the resulting
errors are expected to be of order $e^{-\gamma/\chi}$ or smaller, which is completely negligible in typical quantum optical
systems where $\gamma\gg\chi$.

\subsection{The semi-classical theory}

We can also transcribe the master equation as a c-number phase space evolution equation using the Wigner representation \begin{equation}
P_{W}(\bm\alpha,\bm\alpha^{*})=\frac{1}{\pi^{2}}\int_{-\infty}^{\infty}d^{6}\bm z\;\chi_{W}(\bm z,\bm z^{*})e^{-i\bm z^{*}\cdot\bm\alpha^{*}}e^{-i\bm z\cdot\bm\alpha}\label{a9}\end{equation}
 where $\chi_{W}(\bm z,\bm z^{*})$, the characteristic function for the Wigner representation , is given by \begin{equation}
\chi_{W}(\bm z,\bm z^{*})=Tr\left(\rho e^{i\bm z^{*}\bm a^{\dagger}+i\bm z\cdot\bm a}\right)\label{a8}\end{equation}
 This transcription is particularly useful for semi-classical treatments.

The equation for the Wigner function for the nondegenerate parametric amplifier that corresponds to the master equation (\ref{a2})
turns out to be \begin{eqnarray}
\frac{\partial P_{W}}{\partial t} & = & \left\{ \frac{\partial}{\partial\alpha_{0}}\left(\gamma_{0}\alpha_{0}+\chi\alpha_{1}\alpha_{2}-{\mathcal{E}}\right)\right.\nonumber \\
 &  & +\frac{\partial}{\partial\alpha_{0}^{*}}\left(\gamma_{0}\alpha_{0}^{*}+\chi\alpha_{1}^{*}\alpha_{2}^{*}-{\mathcal{E}}\right)\nonumber \\
 &  & +\frac{\partial}{\partial\alpha_{1}}\left(\gamma_{1}\alpha_{1}-\chi\alpha_{2}^{*}\alpha_{0}\right)+\frac{\partial}{\partial\alpha_{1}^{*}}\left(\gamma_{1}\alpha_{1}^{*}-\chi\alpha_{2}\alpha_{0}^{*}\right)\nonumber \\
 &  & +\frac{\partial}{\partial\alpha_{2}}\left(\gamma_{2}\alpha_{2}-\chi\alpha_{1}^{*}\alpha_{0}\right)+\frac{\partial}{\partial\alpha_{2}^{*}}\left(\gamma_{2}\alpha_{2}^{*}-\chi\alpha_{1}\alpha_{0}^{*}\right)\nonumber \\
 &  & +\gamma_{0}\frac{\partial^{2}}{\partial\alpha_{0}\partial\alpha_{0}^{*}}+\gamma_{1}\frac{\partial^{2}}{\partial\alpha_{1}\partial\alpha_{1}^{*}}+\gamma_{2}\frac{\partial^{2}}{\partial\alpha_{2}\partial\alpha_{2}^{*}}\nonumber \\
 &  & +\left.\frac{\chi}{4}\left(\frac{\partial^{3}}{\partial\alpha_{1}\partial\alpha_{2}\partial\alpha_{0}^{*}}+\frac{\partial^{3}}{\partial\alpha_{1}^{*}\partial\alpha_{2}^{*}\partial\alpha_{0}}\right)\right\} P_{W}\label{a10}\end{eqnarray}

If we drop the third order derivative terms, in an approximation valid in the limit of large photon number, we can equate
the resulting truncated Fokker-Planck equation describing the evolution of the Wigner function with a set of It\^{o} stochastic
equations which read as follows

\begin{eqnarray}
d\alpha_{0} & = & \left({\mathcal{E}}-\gamma_{0}\alpha_{0}-\chi\alpha_{1}\alpha_{2}\right)dt+\sqrt{\gamma_{0}}dW_{0}\nonumber \\
d\alpha_{0}^{*} & = & \left({\mathcal{E}}^{*}-\gamma_{0}\alpha_{0}^{*}-\chi\alpha_{1}^{*}\alpha_{2}^{*}\right)dt+\sqrt{\gamma_{0}}dW_{0}^{*}\nonumber \\
d\alpha_{1} & = & \left(-\gamma_{1}\alpha_{1}+\chi\alpha_{2}^{*}\alpha_{0}\right)dt+\sqrt{\gamma_{1}}dW_{1}\nonumber \\
d\alpha_{1}^{*} & = & \left(-\gamma_{1}\alpha_{1}^{*}+\chi\alpha_{2}\alpha_{0}^{*}\right)dt+\sqrt{\gamma_{1}}dW_{1}^{*}\nonumber \\
d\alpha_{2} & = & \left(-\gamma_{2}\alpha_{2}+\chi\alpha_{1}^{*}\alpha_{0}\right)dt+\sqrt{\gamma_{2}}dW_{2}\nonumber \\
d\alpha_{2}^{*} & = & \left(-\gamma_{2}\alpha_{2}^{*}+\chi\alpha_{1}\alpha_{0}^{*}\right)dt+\sqrt{\gamma_{2}}dW_{2}^{*}\,\,.\label{a11}\end{eqnarray}

\noindent The non-vanishing noise correlations are given by \begin{eqnarray}
\langle dW_{i}\rangle & = & 0\nonumber \\
\langle dW_{i}dW_{i}^{*}\rangle & = & dt~;~i=0,1,2.\label{a12}\end{eqnarray}

If we compare the two sets of It\^{o} stochastic equations, namely (\ref{a6}) and (\ref{a11}), we notice that the main
difference between the two is in the structure of the noise terms. While the noise terms in the positive-P equations (\ref{a6})
depend on the pumping amplitude and the nonlinear coupling constant, those in the Wigner representation do not. In fact they
correspond precisely to the noise terms that one adds, in the linear case, in compliance with the fluctuation-dissipation
theorem.

In some sense, one can interpret the noise terms in the Wigner case as accounting for vacuum fluctuations. However, the truncated
Wigner theory must be treated cautiously, since it neglects important third-order correlations which are not always negligible.
These equations imply that $\langle\alpha_{i}\alpha_{i}^{\dagger}\rangle=\langle\widehat{n}_{i}\rangle=1/2$ when there is
no driving and no coupling, as expected for a vacuum state in a symmetrically-ordered representation. However, a vacuum state
is \emph{not} obtained semi-classically if there is any coupling $\chi$, even with a vacuum input, which is an unphysical
feature. The full Wigner theory has no such limitations: but it is no longer positive-definite, and therefore has no equivalent
stochastic formulation.

\subsection{Comparison of methods}

In comparing these methods, we notice that the classical equations of course give no information at all about the quantum
fluctuations, although they give an excellent guide to the location of the critical point when the threshold photon number
is large.

To include quantum effects, one might imagine that a direct numerical calculation in a photon number basis would be useful,
provided the maximum photon number was small. We note that in a three-mode system, the Hilbert space dimension scales as
$n_{max}^{3}$, while the density matrix has $n_{max}^{6}$components provided the boson number is bounded by $n_{max.}$In
practice, one finds that typical experiments have $n_{max}\simeq10^{3}-10^{9}$. This implies that neither the full density
matrix nor even the reduced wave-function in a stochastic wave-function calculation\cite{Wavefunction} can in general be
calculated directly with current computers, for practical reasons of memory and computational time. Direct number state methods
are also not convenient for analytical approximations.

Other techniques involve Feynman (or related) diagrams, using a hierarchy of correlation functions\cite{Diagram}. These
methods give useful results below threshold, and have similarities to perturbation theory using stochastic methods, which
we discuss later. The drawbacks are that these diagrammatic methods appear less systematic than phase-space methods, since
certain classes of diagrams are discarded, and the results usually diverge at the critical point.

Semi-classical methods overcome some of these limitations. The Wigner truncation approximation does include quantum fluctuation
effects, but is only valid if all photon numbers are large, which is a highly questionable assumption in the signal/idler
modes below threshold. In addition, the neglected third-order derivative terms are at best only suppressed by a polynomial
factor, and there are no other methods to check the accuracy of the approximation. 

By comparison, the approximation of neglecting boundary terms in the positive-P equations appears well-justified in these
calculations as long as $\gamma\gg\chi$. If necessary - that is, if unstable trajectories occur - the technique can be checked
with the more precise stochastic gauge approach. No evidence was found that boundary terms were significant here, even for
the relatively low damping rates used in the numerics. As one might expect, the truncated Wigner method gives rise to unphysical
predictions at low driving field, which does not occur with the positive-P equations - given the parameters used here. Accordingly,
we mainly focus on the positive-P phase-space method in this paper.

\section{Observable moments and EPR spectra }

In order to understand what types of calculation to carry out for this system, it is important to identify observable measurements
that can be carried out, and relate these to operators and their correlations. 

The positive-P stochastic method directly reproduces the normally ordered correlations and moments, while the Wigner representation
reproduces the symmetrically ordered moments. Of course, commutation relations can always be used to transform one type of
ordering into the other. Further, we also have to distinguish between the internal and external operator moments, since measurements
are normally performed on output fields that are external to the cavity. The necessary formalism for treating external field
spectra was introduced and developed by Yurke\cite{Yurke}, and by Collett and Gardiner\cite{Gardiner}.

As we shall see, there is a direct relationship between the output field spectra of a nondegenerate OPO, and observable criteria
for EPR correlations and entanglement.

\subsection{Internal moments}

The squeezing in terms of the intra-cavity quadrature covariances corresponds to an instantaneous measurement of the field
moments\begin{equation}
S_{ij}^{\theta}=\langle:\widehat{X}_{i}^{\theta}\left(t\right)\widehat{X}_{j}^{\theta}\left(t\right):\rangle\,\,,\end{equation}
 where we define\begin{equation}
\widehat{X}_{j}^{\theta}=e^{-i\theta}\widehat{a}_{j}(t)+e^{i\theta}\widehat{a}_{j}^{\dagger}(t)\end{equation}
 to denote internal quadrature operators. Similarly, complex quadratures\cite{Caves} are defined as:

\begin{eqnarray}
\widehat{X}^{\theta} & = & e^{-i\theta}\widehat{a}_{1}(t)+e^{i\theta}\widehat{a}_{2}^{\dagger}(t)\nonumber \\
 & = & \frac{1}{2}\left(\widehat{X}_{1}^{\theta}+\widehat{X}_{2}^{\theta}+i\left(\widehat{X}_{1}^{\theta+\pi/2}-\widehat{X}_{2}^{\theta+\pi/2}\right)\right)\,\,,\end{eqnarray}
with a normally-ordered intra-cavity variance of:\begin{eqnarray}
S^{\theta} & = & \langle:\widehat{X}^{\theta}\left(t\right)\widehat{X}^{\dagger\theta}\left(t\right):\rangle\nonumber \\
 & = & \frac{1}{4}\langle:\left(\widehat{X}_{1}^{\theta}+\widehat{X}_{2}^{\theta}\right)^{2}:\rangle\nonumber \\
 & + & \frac{1}{4}\langle:\left(\widehat{X}_{1}^{\theta+\pi/2}-\widehat{X}_{2}^{\theta+\pi/2}\right)^{2}:\rangle\,\,,\end{eqnarray}

If such measurements were possible, they would include contributions from all frequencies. However, it is more typical that
one has access to spectrally resolved quadrature measurements of the output fields, and these are generally more useful as
measures of entanglement and squeezing.

\subsection{External spectra}

The external field measurements are obtained from the input-output relations : \begin{equation}
\widehat{\Phi}_{j}^{out}(t)=\sqrt{2\gamma_{j}^{out}}\;\widehat{a}_{j}(t)-\widehat{\Phi}_{j}^{in}(t)\,\,,\end{equation}
 where $\widehat{\Phi}_{j}^{in}(t)$ and $\widehat{\Phi}_{j}^{out}(t)$ are the input and output photon fields respectively,
evaluated at the output-coupling mirror, and $\widehat{a}_{j}(t)$ is the intra-cavity photon field. The most efficient transport
of squeezing is obtained if we assume that all the signal losses occur through the output coupler, so that $\gamma_{1}=\gamma_{1}^{out}$.
We will assume this to be the case for simplicity, though the necessary corrections\cite{rd} for imperfect interferometers
simply involve the ratio $\gamma_{j}^{out}/\gamma_{j}$.

The measured output spectral covariance $V_{ij}^{\theta}$ of a general quadrature \begin{equation}
\widehat{X}_{j}^{\theta\; out}=e^{-i\theta}\widehat{\Phi}_{j}^{out}(t)+e^{i\theta}\widehat{\Phi}_{j}^{\dagger out}(t)\end{equation}
 can be written as\begin{equation}
V_{ij}^{\theta}(\omega)\delta(\omega+\omega^{\prime})=\left\langle \Delta\widehat{X}_{i}^{\theta\; out}(\omega)\Delta\widehat{X}_{j}^{\theta\; out}(\omega^{\prime})\right\rangle \,\,,\label{OS-spectra}\end{equation}
 where the fluctuations $\Delta\widehat{X}_{j}^{\theta\; out}$ are defined as $\Delta\widehat{X}_{j}^{\theta\; out}=\widehat{X}_{j}^{\theta\; out}-\langle\widehat{X}_{j}^{\theta\; out}\rangle$,
$\theta$ is a phase angle related to a phase sensitive local oscillator measurement, and the frequency argument denotes
a Fourier transform: \begin{equation}
\widehat{X}_{j}^{\theta\; out}(\omega)=\int\frac{dt}{\sqrt{2\pi}}e^{i\omega t}\widehat{X}_{j}^{\theta\; out}(t)\,\,.\end{equation}

We also introduce complex quadratures and their Fourier transforms, which are useful for computational purposes:\begin{eqnarray}
\widehat{X}^{\theta\; out} & = & e^{-i\theta}\widehat{\Phi}_{1}^{out}(t)+e^{i\theta}\widehat{\Phi}_{2}^{\dagger out}(t)\nonumber \\
\widehat{X}^{\dagger\theta\; out} & = & e^{-i\theta}\widehat{\Phi}_{2}^{out}(t)+e^{i\theta}\widehat{\Phi}_{1}^{\dagger out}(t)\nonumber \\
\widehat{X}^{\theta\; out}(\omega) & = & \int\frac{dt}{\sqrt{2\pi}}e^{i\omega t}\widehat{X}^{\theta\; out}(t)\nonumber \\
\widehat{X}^{\dagger\theta\; out}(\omega) & = & \int\frac{dt}{\sqrt{2\pi}}e^{i\omega t}\widehat{X}^{\dagger\theta\; out}(t)\label{complexquad}\end{eqnarray}
 The spectral quadrature operators $\widehat{X}^{\theta\; out}(\omega)$ are \emph{not} formally hermitian, except at $\omega=0$.

\subsection{Observable quadratures}

In practice, one is mostly interested in external spectral measurements taken over a long but finite interval, after a steady-state
is achieved. For output measurements averaged over a long time $T$, it is the low frequency part of the spectrum that is
the relevant quantity, as it usually determines the maximum squeezing or entanglement possible. For simplicity, we will focus
on the $\omega=0$ case, where we can define \emph{observable} frequency-domain quadrature operators as:\begin{eqnarray}
\widehat{X}_{i}^{o} & = & \sqrt{\frac{2\pi}{T}}\widehat{X}_{i}^{0\; out}(0)\nonumber \\
\widehat{Y}_{i}^{o} & = & \sqrt{\frac{2\pi}{T}}\widehat{X}_{i}^{\pi/2\; out}(0)\,\,.\label{eq:observables}\end{eqnarray}

Since $\delta(\omega)\sim T/2\pi$ when mapping a Fourier transform to a Fourier series, the complex quadrature spectrum
for a finite time interval is normalized as:\begin{equation}
V^{\theta}(0)=\left<\frac{2\pi}{T}\left\langle \widehat{X}^{\theta\; out}(0)\widehat{X}^{\theta\; out}(0)\right\rangle \right>\,\,.\end{equation}
In particular, the most important spectra are the unsqueezed and squeezed spectra defined by:\begin{eqnarray}
V^{0}(0) & = & \frac{1}{4}\left\langle \left[\widehat{X}_{1}^{o}+\widehat{X}_{2}^{o}\right]^{2}+\left[\widehat{Y}_{1}^{o}-\widehat{Y}_{2}^{o}\right]^{2}\right\rangle \nonumber \\
V^{\pi/2}(0) & = & \frac{1}{4}\left\langle \left[\widehat{Y}_{1}^{o}+\widehat{Y}_{2}^{o}\right]^{2}+\left[\widehat{X}_{1}^{o}-\widehat{X}_{2}^{o}\right]^{2}\right\rangle \,\,.\label{eq:Vzero}\end{eqnarray}

In other words, the complex quadrature spectra simply correspond to simultaneous sum and difference measurements on the two
observed output quadratures for the signal and idler, with the precise quadratures observed adjustable via the local oscillator
phase angle $\theta$. 

The properties of external quadratures for $\omega\neq0$ are experimentally important since technical noise normally prohibits
direct quadrature measurements at $\omega=0$. Nevertheless, even at $\omega\neq0$ the quadratures are decomposable\cite{rd}
into pairs of mutually commuting hermitian operators with similar properties to the intra-cavity quadrature operators, by
using discrete sine and cosine transforms. These results therefore hold at non-zero frequencies. 

The correlations are closely related to those proposed by EPR. We will give more details in the next section, explaining
the relationship of this type of measurement with the EPR paradox and entanglement.

\subsection{Stochastic mappings of operator moments}

We now wish to relate these observed operator correlations with the stochastic correlations that are used to calculate them
via the c-number equivalences.

\subsubsection{P-representation}

In the P-representation normally-ordered operator averages directly relate to stochastic moments with respect to the positive
P-function: \begin{equation}
\langle:\widehat{X}_{j}^{\theta}\left(t\right)\widehat{X}_{j}^{\theta}\left(t\right):\rangle=\langle{X}_{j}^{\theta}\left(t\right){X}_{j}^{\theta}\left(t\right)\rangle_{P}\,\,,\end{equation}
where the internal stochastic variables corresponding to the quadratures are denoted:\begin{equation}
X_{j}^{\theta}=\left(\alpha_{j}e^{-i\theta}+\alpha_{j}^{+}e^{i\theta}\right)\,\,.\end{equation}

Also, the positive P- spectral correlations correspond to the normally ordered, time-ordered operator correlations of the
measured fields. This leads to the following well-known result for the general squeezing spectrum, as measured in an external
homodyne detection scheme: \begin{equation}
V_{ij}^{\theta}(\omega)\delta(\omega+\omega^{\prime})=\delta_{ij}+2\sqrt{\gamma_{i}^{out}\gamma_{j}^{out}}\left\langle \Delta\tilde{X}_{i}^{\theta}\left(\omega\right)\Delta\tilde{X}_{j}^{\theta}\left(\omega^{\prime}\right)\right\rangle _{P}\,\,.\end{equation}
This calculation only involves the internal stochastic quadrature spectral variables, defined as:\begin{equation}
\Delta\tilde{X}_{j}^{\theta}\left(\omega\right)=\int\frac{dt}{\sqrt{2\pi}}e^{i\omega t}\left(X_{j}^{\theta}(t)-\langle X_{j}^{\theta}(t)\rangle_{P}\right)\,\,.\end{equation}
 Note that vacuum (input) field terms do not contribute directly to this spectrum, as they have a vanishing normally-ordered
spectrum, and are not correlated with the coherent amplitudes in the positive P- representation.

\subsubsection{Wigner representation}

In the Wigner representation, on the other hand, the moments and correlations with respect to the Wigner function are directly
related to averages of symmetrically ordered operators. It becomes necessary to rewrite the normally-ordered internal field
averages in terms of symmetrically ordered averages using equal-time commutators. As a result the two spectral orderings
are related by:\begin{equation}
\left\langle :\widehat{X}_{i}^{\theta}\left(t\right)\widehat{X}_{j}^{\theta}\left(t\right):\right\rangle =\left\langle {X}_{i}^{\theta}\left(t\right){X}_{j}^{\theta}\left(t\right)\right\rangle _{W}-\delta_{ij}\,\,.\end{equation}

Similarly, for the normally-ordered squeezing spectrum, as measured in an external homodyne detection scheme, one has: \begin{equation}
V_{ij}^{\theta}(\omega)\delta(\omega+\omega^{\prime})=\left\langle \Delta\tilde{X}_{i}^{\theta\; out}\left(\omega\right)\Delta\tilde{X}_{j}^{\theta\; out}\left(\omega^{\prime}\right)\right\rangle _{W}\,\,.\end{equation}
 Here we define Fourier transforms of fluctuations as previously, except with respect to stochastic output fields:\begin{equation}
X_{j}^{\theta\; out}=e^{-i\theta}\Phi_{j}^{out}(t)+e^{i\theta}\Phi_{j}^{\dagger out}(t)\end{equation}
where:\begin{equation}
\Phi_{j}^{out}(t)=\sqrt{2\gamma^{out}}\alpha_{j}-\Phi_{j}^{in}(t)\,\,.\end{equation}

It is essential to include the vacuum field contributions from reflected input fields, as these are correlated with the internal
Wigner amplitudes, and hence contribute significantly to the spectrum. In fact, these input fields can be shown to correspond
directly to the noise terms in the Wigner representation stochastic equations, leading to the identification: \begin{equation}
\frac{dW_{j}}{dt}=\sqrt{2}{\Phi}_{j}^{in}(t)\,\,,\end{equation}
 where ${\Phi}_{j}^{in}(t)$ is a c-number amplitude corresponding (in the Wigner representation) to the quantum vacuum input
field, and $\langle{\Phi}_{j}^{in}(t){\Phi}_{j}^{*in}(t')\rangle_{W}=\delta(t-t')/2$~.

The fundamental property of the Wigner function is that the ensemble average of any polynomial of the random variable $\alpha$
and $\alpha^{*}$ weighted by the Wigner density exactly corresponds to the Hilbert-space expectation of the corresponding
symmetrized product of the annihilation and creation operators. Therefore, the truncated theory with a positive Wigner function
can be viewed as equivalent to a hidden variable theory, since one can obtain quadrature fluctuation predictions by following
an essentially classical prescription; in which even the noise terms have a classical interpretation as corresponding a form
of zero-point fluctuations. This description cannot be equivalent to quantum mechanics in general, but may provide results
which, under some circumstances, turn out to be quite similar to the quantum mechanical results.

\section{EPR-correlations and Entanglement}

A quantitative, experimentally testable criterion for the EPR paradox was proposed in 1989~\cite{eprr}. It is important
to understand the physical interpretation of this paradox. EPR originally assumed local realism, and claimed that an observation
of perfectly correlated positions and momenta would imply the incompleteness of quantum mechanics. A modern interpretation
is that one can merely deduce the \emph{inconsistency} of local realism with quantum mechanical completeness, since local
realism in Einstein's original sense is no longer widely accepted. This is a weaker paradox than the Bell inequality - which
rules out all local realistic interpretations. However, the Bell inequality has not yet been violated, due to causality and/or
measurement inefficiency issues (though weaker inequalities have been violated). The EPR paradox with quadrature variables
has the advantage that the required degree of measurement efficiency is readily achievable with photo-detectors, since it
does not require single-photon counting.

\subsection{1989 EPR criterion: Violation of an Inferred Heisenberg Uncertainty Principle}

Consider two spatially separated subsystems at $A$ and $B$. Observables $\hat{X}_{1}$ (\char`\"{}position\char`\"{})
and $\hat{Y}_{1}$ (\char`\"{}momentum\char`\"{}) are defined for subsystem $A$, where the two operators have a commutator
of $\left[\hat{X}_{1},\hat{Y}_{1}\right]=2$, so that by Heisenberg's uncertainty principle, $\Delta^{2}\hat{X}_{1}\Delta^{2}\hat{Y}_{1}\geq1$.
Suppose that the two subsystems are partially correlated, as may occur in a real experiment, as opposed to the ideal correlations
in the EPR \emph{gedanken}-experiment. One may still predict the result of measurement $\hat{X}_{1}$, based on the result
of a causally separated measurement $\hat{X}_{2}$ performed at $B$. However, the prediction is imperfect, and has an associated
inference error. Also, for a different choice of measurement $\hat{Y}_{2}$ at $B$, suppose that one may predict the result
of measurement $\hat{Y}_{1}$ at $A$.

We define \begin{eqnarray}
\Delta_{inf}^{2}X_{1} & = & \int P(X_{2})\Delta(X_{1}|X_{2})dX_{2}\nonumber \\
\Delta_{inf}^{2}Y_{1} & = & \int P(Y_{2})\Delta(Y_{1}|Y_{2})dY_{2}\end{eqnarray}
 Here $X_{2}$ labels all outcomes of the measurement $\hat{X}_{2}$ at $B$, and $\Delta(X_{1}|X_{2})$ is the standard
deviation of the conditional distribution $P(X_{1}|X_{2})$, where $X_{1}$ is the conditional result of the measurement
$\hat{X}_{1}$ at $A$, given the measurement $\hat{X}_{2}$ at $B$. The probability $P(X_{2})$ is the probability for
a result $X_{2}$ upon measurement of $\hat{X}_{2}$.

Next, we define an inference variance $\Delta_{inf}^{2}\hat{X}_{1}$ as the average variance of the conditional (inference)
variances $\Delta(X_{1}|X_{2})$ for the prediction (inference) of the result $X_{1}$ for $\hat{X}_{1}$ at $A$, conditional
on a measurement $\hat{X}_{2}$ at $B$. We define $\Delta(Y_{1}|Y_{2})$ similarly to represent the weighted variance associated
with the prediction (inference) of the result $\hat{Y}_{1}$ at $A$, based on the result of the measurement at $B$.

The 1989 inferred H.U.P. criterion~\cite{eprr} to demonstrate EPR correlations in the spirit of the EPR paradox is \begin{equation}
\Delta_{inf}X_{1}\Delta_{inf}Y_{1}<1.\label{eqn:eprcrit}\end{equation}

This EPR-style criterion (\ref{eqn:eprcrit}) was not given in the EPR paper, but has the useful property that it represents
a quantitative inequality that can be experimentally satisfied, without having to construct an experimentally impossible
state with perfect correlations, as in the original proposal. As an added advantage, the application of this inequality to
electromagnetic quadrature variables allows the use of efficient photo-detection techniques, which makes this a completely
practical measure.

By contrast, the violation of a Bell inequality - while having stronger consequences - is more difficult to achieve, owing
to poor efficiencies encountered in single-particle detectors and polarizers. For either type of experiment, a crucial element
is the causal separation of detectors. Without this, arguments using causality provide no constraints or inequalities at
all.

\subsubsection{Linear estimate criterion}

It is not always convenient to measure each conditional distribution $P(X_{1}|X_{2})$ and $P(Y_{1}|Y_{2})$ and its associated
mean and variance. A simpler procedure\cite{eprr} is to propose that upon a result $X_{2}$ for the measurement at $B$
the predicted value for the result $X_{1}$ at $A$ is given linearly by the estimate $X_{est}=cX_{2}+d$. The RMS error
in this estimate after optimizing for $d$ is \begin{equation}
\Delta_{inf,L}^{2}\hat{X}_{1}=\langle\delta_{0}^{2}\rangle-\langle\delta_{0}\rangle^{2}.\end{equation}
 where $\delta_{0}=\hat{X}_{1}-c\hat{X}_{2}$. The best choice for $c$ minimizes $\Delta_{inf,L}^{2}\hat{X}$ and can be
adjusted by experiment, or calculated as discussed in~\cite{eprr} to be $c=(\langle\hat{X}_{1},\hat{X}_{2}\rangle)/\Delta^{2}\hat{X}_{2}$
, where we define $\langle\hat{X}_{1},\hat{X}_{2}\rangle=\langle\hat{X}_{1}\hat{X}_{2}\rangle-\langle\hat{X}_{1}\rangle\langle\hat{X}_{2}\rangle$.

Generally the linear estimate will correspond not be the best estimate for the outcome at $A$, based on the result at $B$.
Therefore generally we have $\Delta_{inf,L}\hat{X}\geq\Delta_{inf}\hat{X}$ and $\Delta_{inf,L}\hat{Y}\geq\Delta_{inf}\hat{Y}$\cite{eprr}.
The observation of \begin{equation}
\Delta_{inf,L}\hat{X}_{1}\Delta_{inf,L}\hat{Y}_{1}<1\label{eq:eprlin}\end{equation}
 will then also imply EPR correlations in the spirit of the EPR paradox.

\subsection{An entanglement criterion based on the observation of two-mode squeezing}

Entanglement may be deduced through a whole set of criteria, of which the EPR criterion (\ref{eqn:eprcrit}) is one\cite{eprr}.
It is possible to deduce entanglement through other criteria\cite{content} without the need to prove the strong EPR correlations.
This has significance within quantum mechanics, but not necessarily the broader implications of the EPR criterion. 

Such entanglement criteria, derived by Duan et.al. and Simon\cite{content}, are based on the proof of quantum inseparability,
where the failure of a separable density matrix \begin{equation}
\rho=\sum_{R}P_{R}\rho_{R}^{1}\rho_{R}^{2}\end{equation}
 ($\sum_{R}P_{R}=1$) is proved. Particularly useful for our purposes is a criterion considered by Duan et al, sufficient
to demonstrate entanglement (inseparability). We define \begin{eqnarray}
\delta\hat{X} & = & \hat{X}_{1}-\hat{X}_{2}\nonumber \\
\delta\hat{Y} & = & \hat{Y}_{1}+\hat{Y}_{2}\,\,.\end{eqnarray}
Entanglement is guaranteed provided that the sum of the variances is bounded by: \begin{equation}
\Delta^{2}\delta\hat{X}+\Delta^{2}\delta\hat{Y}<4\,\,.\label{eqn:insepcrit}\end{equation}

This observation of this entanglement criterion (\ref{eqn:insepcrit}) may be identified as a {}``two-mode squeezing''
criterion for entanglement, since the individual criterion \begin{eqnarray}
\Delta^{2}\delta\hat{X}=\langle\{\hat{X}_{1}-\langle\hat{X}_{1}\rangle-(\hat{X}_{2}-\langle\hat{X}_{2}\rangle)\}^{2}\rangle & < & 2\label{eq:onemodesqz}\end{eqnarray}
 is the criterion for the observation of a type of {}``two-mode squeezing''. In this way we see that fields that are two-mode
squeezed with respect to both $X_{1}-X_{2}$ and $Y_{1}+Y_{2}$, each satisfying (\ref{eq:onemodesqz}), are necessarily
entangled.

\subsection{EPR correlations and entanglement of the parametric system}

The EPR correlations are predicted possible for the outputs of the parametric oscillator. For intracavity entanglement, we
define the quadrature phase amplitudes \begin{eqnarray}
\hat{X}_{1} & = & (\hat{a}_{1}+\hat{a}_{1}^{\dagger})\nonumber \\
\hat{Y}_{1} & = & (\hat{a}_{1}-\hat{a}_{1}^{\dagger})/i\nonumber \\
\hat{X}_{2} & = & (\hat{a}_{2}+\hat{a}_{2}^{\dagger})\nonumber \\
\hat{Y}_{2} & = & (\hat{a}_{2}-\hat{a}_{2}^{\dagger})/i.\label{eqn:quad}\end{eqnarray}
 and identify correlated observables for the oscillator, so that $X_{1}$ is correlated with $X_{2}$ and $Y_{1}$ is correlated
with $-Y_{2}$. The Heisenberg uncertainty relation for the orthogonal amplitudes of mode $\hat{a}_{1}$ is $\Delta^{2}X_{1}\Delta^{2}Y_{1}\geq1$. 

As explained in the previous section, for practical reasons it is preferable to use the corresponding observable external
quadratures defined at or near zero frequency, which are $\widehat{X}_{i}^{o},\widehat{Y}_{i}^{o}$. However, the detailed
arguments only depend on having the commutators defined above, together with the requirement of causality - that is, the
observations must take place with space-like separations between the two detectors over the whole observation period $T$. 

We calculate several types of EPR or entanglement measures. Firstly we evaluate the the 1989 inferred H. U. P. EPR criterion
(\ref{eqn:eprcrit}) but using the linear estimate form, which will allow demonstration of both entanglement and EPR correlations
defined in the spirit of the original EPR paradox. In terms of quadrature phase amplitude measurements this strong EPR criterion
is satisfied when \begin{equation}
\Delta_{inf,L}^{2}X^{o}\Delta_{inf,L}^{2}Y^{o}=\Delta^{2}(X_{1}^{o}-c_{x}X_{2}^{o})\Delta^{2}(Y_{1}^{o}-c_{y}Y_{2}^{o})<1\label{eqn:strong}\end{equation}
 Now $c_{x}=\langle X_{1}^{o},X_{2}^{o}\rangle/\Delta^{2}X_{2}^{o}$ and $c_{y}=\langle Y_{1}^{o},Y_{2}^{o}\rangle/\Delta^{2}Y_{2}^{o}$
will minimize\cite{eprr} the inference variances. Substituting for $c_{x}$ and $c_{y}$, we explicitly calculate \begin{equation}
\Delta_{inf,L}^{2}X^{o}=\Delta^{2}X_{1}^{o}-\langle X_{1}^{o},X_{2}^{o}\rangle^{2}/\Delta^{2}X_{2}^{o}\end{equation}
 and \begin{equation}
\Delta_{inf,L}^{2}Y^{o}=\Delta^{2}Y_{1}^{o}-\langle Y_{1}^{o},Y_{2}^{o}\rangle^{2}/\Delta^{2}Y_{2}^{o}\,\,.\end{equation}
 For our particular system moments such as have $\langle a_{1}\rangle=\langle a_{2}\rangle=...$ are zero and we have symmetry
between $a_{1}$and $a_{2}$ modes, so that \begin{equation}
\Delta^{2}X_{1}^{o}=\frac{1}{2}\left(V+V^{\pi/2}\right)\geq1\end{equation}
 and\begin{equation}
\langle X_{1}^{o},X_{2}^{o}\rangle=\frac{1}{2}\left(V^{0}-V^{\pi/2}\right)\,\,.\end{equation}
 The linear inference EPR criterion (\ref{eq:eprlin}) is then equivalent to:\begin{equation}
\Delta_{inf,L}^{2}X^{o}=\frac{2V^{0}V^{\pi/2}}{V^{0}+V^{\pi/2}}<1\,\,.\label{eq:infvariance}\end{equation}

This criterion is not equivalent to (\ref{eqn:eprcrit}) which is based on the conditionals, since the linear estimate may
not be the best, in which case it is possible that (\ref{eqn:eprcrit}) is satisfied while (\ref{eqn:strong}) is not, and
we do not pick up EPR and entanglement where it exists. Nevertheless the criterion (\ref{eq:infvariance}) is sufficient
to prove EPR correlations and entanglement.

Secondly, we calculate the Duan and Simon et. al. two-mode squeezing criteria (\ref{eqn:insepcrit}) for entanglement. Written
in terms of quadrature phase amplitude measurements this becomes \begin{equation}
V^{\pi/2}=\frac{1}{4}\left(\Delta^{2}(X_{1}^{o}-X_{2}^{o})+\Delta^{2}(Y_{1}^{o}+Y_{2}^{o})\right)<1.\label{eq:twomodequad}\end{equation}
 This criterion was explicitly shown to be both sufficient and necessary for entanglement for the case of Gaussian states
(for appropriately chosen quadratures), meaning that in this case it would pick up any entanglement present. Our system is
not Gaussian, and while these criteria are still sufficient to imply entanglement, they may not be necessary.

It is always the case that for ideal squeezing, both the linear EPR and the squeezed entanglement criteria are satisfied.
Where one has additional loss, however, it is possible for the squeezed-entanglement criterion (\ref{eq:twomodequad}) to
be satisfied but not the EPR criterion (\ref{eqn:strong}). Such situations have been studied by Bowen et al\cite{bow}.
Our situation could be different again, due to the fact that the underlying quantum states undergo nonlinear fluctuations
and are inherently non-Gaussian.

\section{Below-Threshold Intra-cavity moments}

In this section we use perturbation methods to study the nondegenerate parametric oscillator beyond the linearized regime
both in the fully quantum mechanical approach using positive-P representation, and in the semi-classical approach based on
the Wigner function. In the positive P case the basic quantities investigated are correlations involving the internal quadrature
operators, mapped into stochastic variables according to\begin{eqnarray}
X_{0}=\left(\alpha_{0}+\alpha_{0}^{+}\right) &  & \;\;\;\;\;\;\;\;\; Y_{0}=\frac{1}{i}\left(\alpha_{0}-\alpha_{0}^{+}\right)\nonumber \\
X=\left(\alpha_{1}+\alpha_{2}^{+}\right) &  & \;\;\;\;\;\;\;\;\; Y=\frac{1}{i}\left(\alpha_{1}-\alpha_{2}^{+}\right)\nonumber \\
X^{+}=\left(\alpha_{2}+\alpha_{1}^{+}\right) &  & \;\;\;\;\;\;\;\;\; Y^{+}=\frac{1}{i}\left(\alpha_{2}-\alpha_{1}^{+}\right)\label{quad}\end{eqnarray}

In the truncated Wigner (semi-classical) case, we have a similar set with $\alpha_{i}^{+}$ replaced by $\alpha_{i}^{\ast}$.
To avoid excessive notation we use the same symbols for the quadrature variables in the two cases, noting that in the semi-classical
case, $X^{+}=X^{\ast}$ and $Y^{+}=Y^{\ast}$.

For developing a systematic perturbation procedure, it proves convenient to define \begin{equation}
\gamma_{r}=\gamma_{0}/\gamma~~~,~~~\mu=\mathcal{\mathcal{E}}/\mathcal{\mathcal{E}}_{c},~~~g=\frac{\chi}{\gamma\sqrt{2\gamma_{r}}}\label{a17}\end{equation}
 and to introduce the following scaled quadrature variables \begin{eqnarray}
x_{0} & = & g\sqrt{2\gamma_{r}}X_{0}\nonumber \\
y_{0} & = & g\sqrt{2\gamma_{r}}Y_{0}\nonumber \\
x & = & gX\nonumber \\
y & = & gY\nonumber \\
x^{+} & = & gX^{+}\label{a16}\\
y^{+} & = & gY^{+}\,\,.\nonumber \end{eqnarray}

In terms of these new variables, and a scaled time $\tau=\gamma t$, the equations for the quadratures become

\begin{itemize}
\item Positive-P equations \begin{eqnarray}
dx_{0} & = & -\gamma_{r}\left[x_{0}-2\mu+\left(xx^{+}-yy^{+}\right)\right]d\tau\nonumber \\
dy_{0} & = & -\gamma_{r}\left[y_{0}+\left(xy^{+}+yx^{+}\right)\right]d\tau\nonumber \\
dx & = & \left[-x+\frac{1}{2}\left(xx_{0}+yy_{0}\right)\right]d\tau+\frac{g}{\sqrt{2}}\left[\sqrt{x_{0}+iy_{0}}dw_{1}+\sqrt{x_{0}-iy_{0}}dw_{2}^{+}\right]\nonumber \\
dy & = & \left[-y+\frac{1}{2}\left(xy_{0}-yx_{0}\right)\right]d\tau-i\frac{g}{\sqrt{2}}\left[\sqrt{x_{0}+iy_{0}}dw_{1}-\sqrt{x_{0}-iy_{0}}dw_{2}^{+}\right]\nonumber \\
dx^{+} & = & \left[-x^{+}+\frac{1}{2}\left(x^{+}x_{0}+y^{+}y_{0}\right)\right]d\tau+\frac{g}{\sqrt{2}}\left[\sqrt{x_{0}+iy_{0}}dw_{2}+\sqrt{x_{0}-iy_{0}}dw_{1}^{+}\right]\nonumber \\
dy^{+} & = & \left[-y^{+}+\frac{1}{2}\left(x^{+}y_{0}-y^{+}x_{0}\right)\right]d\tau-i\frac{g}{\sqrt{2}}\left[\sqrt{x_{0}+iy_{0}}dw_{2}-\sqrt{x_{0}-iy_{0}}dw_{1}^{+}\right]\label{a18}\end{eqnarray}
\\
where: $\langle dw_{1}dw_{2}\rangle=\langle dw_{1}^{+}dw_{2}^{+}\rangle=d\tau\,\,.$
\item Semi-classical equations 
\end{itemize}
\begin{eqnarray}
dx_{0} & = & -\gamma_{r}\left[x_{0}-2\mu+\left(xx^{+}-yy^{+}\right)\right]d\tau+\sqrt{2}g\gamma_{r}\left[dw_{0}+dw_{0}^{*}\right]\nonumber \\
dy_{0} & = & -\gamma_{r}\left[y_{0}+\left(xy^{+}+yx^{+}\right)\right]d\tau-i\sqrt{2}g\gamma_{r}\left[dw_{0}-dw_{0}^{*}\right]\nonumber \\
dx & = & \left[-x+\frac{1}{2}\left(xx_{0}+yy_{0}\right)\right]d\tau+g\left[dw_{1}+dw_{2}^{*}\right]\nonumber \\
dy & = & \left[-y+\frac{1}{2}\left(xy_{0}-yx_{0}\right)\right]d\tau-ig\left[dw_{1}-dw_{2}^{*}\right]\nonumber \\
dx^{+} & = & \left[-x^{+}+\frac{1}{2}\left(x^{+}x_{0}+y^{+}y_{0}\right)\right]d\tau+g\left[dw_{2}+dw_{1}^{*}\right]\nonumber \\
dy^{+} & = & \left[-y^{+}+\frac{1}{2}\left(x^{+}y_{0}-y^{+}x_{0}\right)\right]d\tau-ig\left[dw_{2}-dw_{1}^{*}\right]\,\,.\label{a19}\end{eqnarray}
\\
where: $\langle dw_{i}dw_{j}^{*}\rangle=\delta_{ij}d\tau\,\,.$

In order to solve these coupled equations systematically, we introduce a formal perturbation expansion in powers of $g$\begin{eqnarray}
x_{k} & = & \sum_{n=0}^{\infty}g^{n}x_{k}^{(n)}\nonumber \\
y_{k} & = & \sum_{n=0}^{\infty}g^{n}y_{k}^{(n)}\label{a20}\end{eqnarray}

This expansion has the property that the zero-th order term corresponds to the large classical field of order $1/g$ in the
unscaled quadratures, while the first order term involves the quantum fluctuation of order $1$, and the higher order terms
correspond to nonlinear corrections to the quantum fluctuations of order $g$ and higher.

\subsection{Matched power equations in Positive-P representation}

Substituting (\ref{a20}) in (\ref{a18}) and equating like powers of $g$ on both sides we obtain a hierarchy of stochastic
equations. The set of equations thus obtained, if desired, can be diagrammatically analyzed using the 'stochastic diagram'
method\cite{Stochdiagram}. The zero-th order equations are \begin{eqnarray}
dx_{0}^{(0)} & = & -\gamma_{r}\left[x_{0}^{(0)}-2\mu+\left(x^{(0)}x^{+(0)}-y^{(0)}y^{+(0)}\right)\right]d\tau\nonumber \\
dy_{0}^{(0)} & = & -\gamma_{r}\left[y_{0}^{(0)}+\left(x^{(0)}y^{+(0)}+y^{(0)}x^{+(0)}\right)\right]d\tau\nonumber \\
dx^{(0)} & = & \left[-x^{(0)}+\frac{1}{2}\left(x^{(0)}x_{0}^{(0)}+y^{(0)}y_{0}^{(0)}\right)\right]d\tau\nonumber \\
dy^{(0)} & = & \left[-y^{(0)}+\frac{1}{2}\left(x^{(0)}y_{0}^{(0)}-y^{(0)}x_{0}^{(0)}\right)\right]d\tau\nonumber \\
dx^{+(0)} & = & \left[-x^{+(0)}+\frac{1}{2}\left(x^{+(0)}x_{0}^{(0)}+y^{+(0)}y_{0}^{(0)}\right)\right]d\tau\nonumber \\
dy^{+(0)} & = & \left[-y^{+(0)}+\frac{1}{2}\left(x^{+(0)}y_{0}^{(0)}-y^{+(0)}x_{0}^{(0)}\right)\right]d\tau\label{a21}\end{eqnarray}

These equations correspond to the classical nonlinear equations for the intra-cavity quadratures expressed in terms of dimensionless
scaled fields. Below threshold, the steady-state solution of these equations is well known and is given by: \begin{eqnarray}
x_{0}^{(0)} & = & 2\mu\label{a22}\\
y_{0}^{(0)} & = & x^{(0)}=y^{(0)}=0\nonumber \end{eqnarray}

\noindent The first order equations are \begin{eqnarray}
dx_{0}^{(1)} & = & -\gamma_{r}x_{0}^{(1)}d\tau\nonumber \\
dy_{0}^{(1)} & = & -\gamma_{r}y_{0}^{(1)}d\tau\nonumber \\
dx^{(1)} & = & -\left(1-\mu\right)x^{(1)}d\tau+\sqrt{2\mu}dw_{x1}\nonumber \\
dy^{(1)} & = & -\left(1+\mu\right)y^{(1)}d\tau-i\sqrt{2\mu}dw_{y1}\nonumber \\
dx^{+(1)} & = & -\left(1-\mu\right)x^{+(1)}d\tau+\sqrt{2\mu}dw_{x2}\nonumber \\
dy^{+(1)} & = & -\left(1+\mu\right)y^{+(1)}d\tau-i\sqrt{2\mu}dw_{y2}\,\,.\label{a23}\end{eqnarray}
 We have introduced new Wiener increments as $dw_{x1(y1)}(\tau)=(dw_{1}(\tau)\pm dw_{2}^{+}(\tau))/\sqrt{2}$ and $dw_{x2(y2)}(\tau)=(dw_{2}(\tau)\pm dw_{1}^{+}(\tau))/\sqrt{2}$,
with the following correlations \begin{eqnarray}
\langle dw_{x1}dw_{x2}\rangle & = & d\tau\nonumber \\
\langle dw_{y1}dw_{y2}\rangle & = & d\tau\,\,.\label{a24}\end{eqnarray}
 and all other correlations vanishing.

The equations (\ref{a23}) are the ones that are normally used to predict squeezing. They are linear stochastic equations
with non classical Gaussian white noise and, if higher-order corrections are ignored, yield an ideal squeezed state for the
sub-harmonic quadratures together with an ideal coherent state for the pump. Further, from the structure of these equations,
it is evident that the steady state solution for the pump field quadratures, in this order, vanish. We can, therefore, without
loss of generality, set all odd orders of $x_{0}^{(n)}$, $y_{0}^{(n)}$ for the pump, and all even orders of $x_{i}^{(n)}$,
$y_{i}^{(n)}~~;~~i=1,2$ for the signal and idler fields respectively equal to zero. With this in mind, the second order
equations turn out to be \begin{eqnarray}
dx_{0}^{(2)} & = & -\gamma_{r}\left[x_{0}^{(2)}+x^{(1)}x^{+(1)}-y^{(1)}y^{+(1)}\right]d\tau\nonumber \\
dy_{0}^{(2)} & = & -\gamma_{r}\left[y_{0}^{(2)}+x^{(1)}y^{+(1)}+y^{(1)}x^{+(1)}\right]d\tau\,\,.\label{a25}\end{eqnarray}

Since, in the present work, our primary interest is to calculate the first nonlinear corrections to ideal squeezed-state
behavior, to be consistent, we need to include contributions from the third order equations as well. These equations are
as given below

\begin{eqnarray}
dx^{(3)} & = & \left[-\left(1-\mu\right)x^{(3)}+\frac{1}{2}\left(x^{(1)}x_{0}^{(2)}+y^{(1)}y_{0}^{(2)}\right)\right]d\tau+\frac{1}{2\sqrt{2\mu}}\left[x_{0}^{(2)}dw_{x1}+iy_{0}^{(2)}dw_{y1}\right]\nonumber \\
dy^{(3)} & = & \left[-\left(1+\mu\right)y^{(3)}+\frac{1}{2}\left(x^{(1)}y_{0}^{(2)}-x_{0}^{(2)}y^{(1)}\right)\right]d\tau+\frac{1}{2\sqrt{2\mu}}\left[y_{0}^{(2)}dw_{x1}-ix_{0}^{(2)}dw_{y1}\right]\nonumber \\
dx^{+(3)} & = & \left[-\left(1-\mu\right)x^{+(3)}+\frac{1}{2}\left(x^{+(1)}x_{0}^{(2)}+y^{+(1)}y_{0}^{(2)}\right)\right]d\tau+\frac{1}{2\sqrt{2\mu}}\left[x_{0}^{(2)}dw_{x2}+iy_{0}^{(2)}dw_{y2}\right]\nonumber \\
dy^{+(3)} & = & \left[-\left(1+\mu\right)y^{+(3)}+\frac{1}{2}\left(x^{+(1)}y_{0}^{(2)}-x_{0}^{(2)}y^{+(1)}\right)\right]d\tau+\frac{1}{2\sqrt{2\mu}}\left[y_{0}^{(2)}dw_{x2}-ix_{0}^{(2)}dw_{y2}\right]\,\,.\label{a26}\end{eqnarray}
 This set of equations has non-trivial noise terms as they depend on the solutions of the stochastic equations at the second
order.

\subsection{Operator moments in the positive P-representation}

The set of stochastic equations equations together with the It\^{o} rules for variable changes\cite{Arnold} permit computation
of the operator moments in a straightforward manner. Apart from their intrinsic interest, they are useful in checking the
correctness of somewhat more involved spectral calculations given later. The results obtained are summarized below: \begin{eqnarray}
\langle x_{0}^{(2)}\rangle & = & \frac{-2\mu^{2}}{1-\mu^{2}}\nonumber \\
\langle y^{(1)}y^{+(1)}\rangle & = & -\left(\frac{\mu}{1+\mu}\right)\nonumber \\
\langle x^{(1)}x^{+(1)}\rangle & = & \left(\frac{\mu}{1-\mu}\right)\nonumber \\
\langle y^{(1)}y^{+(3)}\rangle & = & \frac{\mu}{4\left(1+\mu\right)\left(1-\mu^{2}\right)}\nonumber \\
 & \times & \left[\frac{\mu\gamma_{r}}{\gamma_{r}+2}+\frac{\gamma_{r}\left(2-\mu+\mu^{2}\right)+4\left(1+\mu\right)}{\left(1+\mu\right)\left(\gamma_{r}+2\left(1+\mu\right)\right)}\right]\nonumber \\
\langle x^{(1)}y^{+(1)}y_{0}^{(2)}\rangle & = & \frac{\mu^{2}}{1-\mu^{2}}\left(\frac{\gamma_{r}}{\gamma_{r}+2}\right)\label{a34}\end{eqnarray}
 The first quantity above pertains to the depletion of the pump that supplies energy for the subharmonic mode. The next two
quantities are the squeezed and enhanced quadratures as given by the linearized theory, while the fourth one is the first
correction to the linearized theory. The last one is the steady state triple quadrature correlation. This quantity has been
investigated earlier for its relevance in distinguishing quantum mechanics from a local hidden variable theory\cite{Triple}.

The results above yield the following expression for the steady state intra-cavity squeezed quadrature fluctuations: \begin{eqnarray}
\left\langle \hat{Y}_{1}\hat{Y}_{1}^{\dagger}\right\rangle _{ss} & = & 1+\left\langle :\hat{Y}_{1}\hat{Y}_{1}^{\dagger}:\right\rangle \nonumber \\
 & = & \frac{1}{1+\mu}+\frac{g^{2}\mu}{2\left(1+\mu\right)\left(1-\mu^{2}\right)}\left[\frac{\mu\gamma_{r}}{\gamma_{r}+2}+\frac{\gamma_{r}\left(2-\mu+\mu^{2}\right)+4\left(1+\mu\right)}{\left(1+\mu\right)\left(\gamma_{r}+2\left(1+\mu\right)\right)}\right]\,\,.\end{eqnarray}

\begin{figure}
\includegraphics[%
  width=8cm]{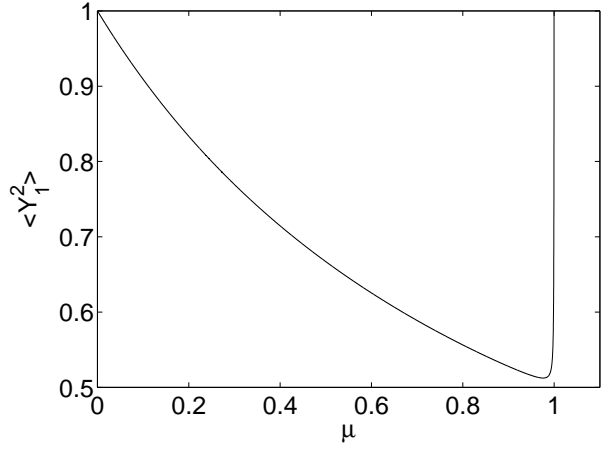}

\caption{Graph of total squeezing/entanglement moment $\langle\hat{Y}_{1}^{2}\rangle$ vs driving field $\mu$, using parameters
of $g^{2}=0.001$, and $\gamma_{r}=0.5$. This demonstrates the existence of intra-cavity squeezing and entanglement for
$\mu<1$.\label{Graph-of-totalnlmom}}
\end{figure}

Note that the intracavity squeezing quadrature near threshold is not perfectly squeezed, having a limiting squeezing/entanglement
of $0.5$, as shown in Fig (\ref{Graph-of-totalnlmom}).

As might be expected, the nonlinear correction is divergent at the threshold, and needs to be handled either by numerical
integration or a critical-point expansion. Questions relating to optimal output entanglement and squeezing will be treated
in the next section, using frequency domain methods.

\subsection{Matched power equations in semi-classical theory}

Using the same technique of matching the powers of $g$, we obtain the following set of equations in the semi-classical theory.
The zero-th order equation are: \begin{eqnarray}
dx_{0}^{(0)} & = & -\gamma_{r}\left[x_{0}^{(0)}-2\mu+\left(x^{(0)}x^{+(0)}-y^{(0)}y^{+(0)}\right)\right]d\tau\nonumber \\
dy_{0}^{(0)} & = & -\gamma_{r}\left[y_{0}^{(0)}+\left(x^{(0)}y^{+(0)}+y^{(0)}x^{+(0)}\right)\right]d\tau\nonumber \\
dx^{(0)} & = & \left[-x^{(0)}+\frac{1}{2}\left(x^{(0)}x_{0}^{(0)}+y^{(0)}y_{0}^{(0)}\right)\right]d\tau\nonumber \\
dy^{(0)} & = & \left[-y^{(0)}+\frac{1}{2}\left(x^{(0)}y_{0}^{(0)}-y^{(0)}x_{0}^{(0)}\right)\right]d\tau\nonumber \\
dx^{+(0)} & = & \left[-x^{+(0)}+\frac{1}{2}\left(x^{+(0)}x_{0}^{(0)}+y^{+(0)}y_{0}^{(0)}\right)\right]d\tau\nonumber \\
dy^{+(0)} & = & \left[-y^{+(0)}+\frac{1}{2}\left(x^{+(0)}y_{0}^{(0)}-y^{+(0)}x_{0}^{(0)}\right)\right]d\tau\,\,.\label{a27}\end{eqnarray}

As in the positive-P case, the steady-state solution of these equations is given by: \begin{eqnarray}
x_{0}^{(0)} & = & 2\mu\label{a28}\\
y_{0}^{(0)} & = & x^{(0)}=y^{(0)}=0\,\,.\nonumber \end{eqnarray}

The first order equations are \begin{eqnarray}
dx_{0}^{(1)} & = & -\gamma_{r}x_{0}^{(1)}d\tau+2\gamma_{r}dw_{x0}\nonumber \\
dy_{0}^{(1)} & = & -\gamma_{r}y_{0}^{(1)}d\tau+2\gamma_{r}dw_{y0}\nonumber \\
dx^{(1)} & = & -\left(1-\mu\right)x^{(1)}d\tau+\sqrt{2}dw_{x1}\nonumber \\
dy^{(1)} & = & -\left(1+\mu\right)y^{(1)}d\tau+\sqrt{2}dw_{y1}\nonumber \\
dx^{+(1)} & = & -\left(1-\mu\right)x^{+(1)}d\tau+\sqrt{2}dw_{x2}\nonumber \\
dy^{+(1)} & = & -\left(1+\mu\right)y^{+(1)}d\tau+\sqrt{2}dw_{y2}\,\,,\label{a29}\end{eqnarray}

\noindent where \begin{equation}
\langle dw_{x0}dw_{x0}\rangle=\langle dw_{y0}dw_{y0}\rangle=\langle dw_{x1}dw_{x2}\rangle=\langle dw_{y1}dw_{y2}\rangle=d\tau\,\,,\label{a30}\end{equation}
 with all other correlations vanishing.

The equations above give the linearized theory. The first nonlinear corrections come from the next two sets of equations
given below.

The second order equations are: \begin{eqnarray}
dx_{0}^{(2)} & = & -\gamma_{r}\left[x_{0}^{(2)}+x^{(1)}x^{+(1)}-y^{(1)}y^{+(1)}\right]d\tau\nonumber \\
dy_{0}^{(2)} & = & -\gamma_{r}\left[y_{0}^{(2)}+x^{(1)}y^{+(1)}+y^{(1)}x^{+(1)}\right]d\tau\nonumber \\
dx^{(2)} & = & \left[-\left(1-\mu\right)x^{(2)}+\frac{1}{2}\left(x^{(1)}x_{0}^{(1)}+y^{(1)}y_{0}^{(1)}\right)\right]d\tau\nonumber \\
dy^{(2)} & = & \left[-\left(1+\mu\right)y^{(2)}+\frac{1}{2}\left(x^{(1)}y_{0}^{(1)}-x_{0}^{(1)}y^{(1)}\right)\right]d\tau\nonumber \\
dx^{+(2)} & = & \left[-\left(1-\mu\right)x^{+(2)}+\frac{1}{2}\left(y^{+(1)}y_{0}^{(1)}+x^{+(1)}x_{0}^{(1)}\right)\right]d\tau\nonumber \\
dy^{+(2)} & = & \left[-\left(1+\mu\right)y^{+(2)}+\frac{1}{2}\left(x^{+(1)}y_{0}^{(1)}-x_{0}^{(1)}y^{+(1)}\right)\right]d\tau\,\,.\label{a31}\end{eqnarray}

The third order equations are: \begin{eqnarray}
dx_{0}^{(3)} & = & -\gamma_{r}\left[x_{0}^{(3)}+x^{(1)}x^{+(2)}+x^{(2)}x^{+(1)}-y^{(1)}y^{+(2)}-y^{(2)}y^{+(1)}\right]d\tau\nonumber \\
dy_{0}^{(3)} & = & -\gamma_{r}\left[y_{0}^{(3)}+x^{(1)}y^{+(2)}+x^{(2)}y^{+(1)}+y^{(1)}x^{+(2)}+y^{(2)}x^{+(1)}\right]d\tau\nonumber \\
dx^{(3)} & = & \left[-\left(1-\mu\right)x^{(3)}+\frac{1}{2}\left(x^{(1)}x_{0}^{(2)}+x^{(2)}x_{0}^{(1)}+y^{(1)}y_{0}^{(2)}+y^{(2)}y_{0}^{(1)}\right)\right]d\tau\nonumber \\
dy^{(3)} & = & \left[-\left(1+\mu\right)y^{(3)}+\frac{1}{2}\left(x^{(1)}y_{0}^{(2)}+x^{(2)}y_{0}^{(1)}-y^{(1)}x_{0}^{(2)}-y^{(2)}x_{0}^{(1)}\right)\right]d\tau\nonumber \\
dx^{+(3)} & = & \left[-\left(1-\mu\right)x^{+(3)}+\frac{1}{2}\left(x^{+(1)}x_{0}^{(2)}+x^{+(2)}x_{0}^{(1)}+y^{+(1)}y_{0}^{(2)}+y^{+(2)}y_{0}^{(1)}\right)\right]d\tau\nonumber \\
dy^{+(3)} & = & \left[-\left(1+\mu\right)y^{+(3)}+\frac{1}{2}\left(x^{+(1)}y_{0}^{(2)}+x^{+(2)}y_{0}^{(1)}-y^{+(1)}x_{0}^{(2)}-y^{+(2)}x_{0}^{(1)}\right)\right]d\tau\,\,.\label{a32}\end{eqnarray}

\subsection{Operator moments in semi-classical theory}

In this case, the analogues of the results in (\ref{a34}) are found to be: \begin{eqnarray}
\langle x_{0}^{(2)}\rangle & = & \frac{-2\mu^{2}}{1-\mu^{2}}\nonumber \\
\langle x^{(1)}x^{+(1)}\rangle & = & \left(\frac{1}{1-\mu}\right)\nonumber \\
\langle y^{(1)}y^{+(1)}\rangle & = & \left(\frac{1}{1+\mu}\right)\nonumber \\
\langle y^{(2)}y^{+(2)}\rangle & = & \frac{1}{2(1-\mu)(1+\mu)}\left(\frac{\gamma_{r}}{\gamma_{r}+2}\right)+\frac{1}{2(1+\mu)^{2}}\left(\frac{\gamma_{r}}{\gamma_{r}+2(1+\mu)}\right)\nonumber \\
\langle y^{(1)}y^{+(3)}\rangle & = & -\frac{\mu}{4(1-\mu)(1+\mu)^{2}}\left(\frac{\gamma_{r}}{\gamma_{r}+2}\right)+\frac{\mu}{2(1-\mu)(1+\mu)^{3}}+\frac{\mu}{4(1+\mu)^{3}}\left[\frac{\gamma_{r}}{\gamma_{r}+2(1+\mu)}\right]\nonumber \\
\langle x^{(1)}y^{+(1)}y_{0}^{(2)}\rangle & + & \langle x^{(2)}y^{+(1)}y_{0}^{(1)}\rangle+\langle x^{(1)}y^{+(2)}y_{0}^{(1)}\rangle=\frac{1}{1-\mu^{2}}\left(\frac{\gamma_{r}}{\gamma_{r}+2}\right)\,\,.\label{a33}\end{eqnarray}

The main difference in these calculation, compared with the positive-P results, appears in the nonlinear correction for the
subharmonic squeezed quadrature. Up to second order in $g$ we have \begin{eqnarray}
\langle\hat{Y}_{1}^{2}\rangle & = & \frac{1}{g^{2}}\left[{g^{2}}\langle y^{(1)}y^{(1)}\rangle+{g^{4}}\langle y^{(2)}y^{(2)}\rangle+2{g^{4}}\langle y^{(1)}y^{(3)}\rangle\right]\nonumber \\
 & = & \frac{1}{1+\mu}+\frac{g^{2}}{2(1+\mu)(1-\mu^{2})}\left[\frac{\gamma_{r}}{\gamma_{r}+2}+\frac{\gamma_{r}(1+3\mu-2\mu^{2})+4\mu(1+\mu)}{(1+\mu)[\gamma_{r}+2(1+\mu)]}\right]\,\,.\end{eqnarray}

The similarities and disagreement between this result and the positive-P expression for the same quantity deserve further
comments given in the concluding section. In particular, we note that, while the linear terms agree, the nonlinear term are
not in agreement below threshold. However, just below threshold both theories give essentially identical nonlinear corrections.
There is good agreement also in the limit $\gamma_{r}\rightarrow0$.

\begin{figure}
\includegraphics[%
  clip,
  width=8cm]{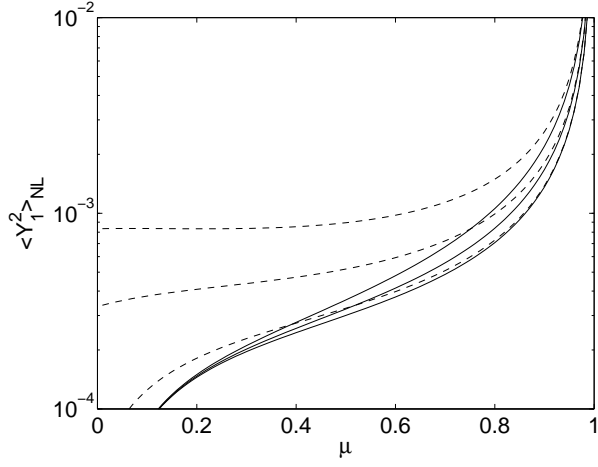}

\caption{Graph of second order nonlinear correction to the squeezing/entanglement moment $\langle\hat{Y}_{1}^{2}\rangle_{NL}$ vs
driving field $\mu$, using parameters of $g^{2}=0.001$, and $\gamma_{r}=0.1,1,10$. Upper lines have larger $\gamma_{r}$
values. Solid lines are the positive-P result, which vanish at small driving field. Dotted lines are the (less accurate)
semi-classical result, which do not vanish at small driving field. \label{Graph-of-nlmom}}
\end{figure}

These comparisons are shown in Fig (\ref{Graph-of-nlmom}).

\section{Spectral Correlations}

Next, we proceed to analyze spectral correlations which are of direct relevance to comparison with experiments. In particular,
we compute the nonlinear corrections to the squeezing spectrum.

\subsection{Positive-P representation}

To perform calculations in the frequency domain, it proves convenient to deal directly with the Fourier transforms \[
\tilde{x}\left(\Omega\right)=\int\frac{d\tau}{\sqrt{2\pi}}e^{i\Omega\tau}x\left(\tau\right)\]
of the hierarchy of the stochastic equations obtained earlier. The equations thus obtained contain noise terms \[
\xi_{x,y}\left(\Omega\right)=\int\frac{d\tau}{\sqrt{2\pi}}e^{i\Omega\tau}\xi_{x,y}\left(\tau\right)\]
 with the following correlations: \begin{eqnarray}
\left\langle \xi_{a}\left(\Omega\right)\right\rangle  & = & 0\,\,,\nonumber \\
\left\langle \xi_{a1}\left(\Omega\right)\xi_{b2}\left(\Omega^{\prime}\right)\right\rangle  & = & \delta_{ab}\delta\left(\Omega+\Omega^{\prime}\right)\,\,.\end{eqnarray}
 In this context, for notational compactness it is useful to introduce the standard notation for convolution of two functions:

\[
[A\star B](\Omega)=\int\frac{d\Omega^{\prime}}{\sqrt{2\pi}}A(\Omega^{\prime})B(\Omega-\Omega^{\prime})\,\,.\]
 With this in mind, the stochastic equations obtained earlier may be rewritten in the frequency domain as follows:

\begin{itemize}
\item First order: \begin{eqnarray}
{\tilde{x}}^{\left(1\right)}\left(\Omega\right) & = & \frac{\sqrt{2\mu}\;\xi_{x1}\left(\Omega\right)}{\left(-i\Omega+1-\mu\right)}\nonumber \\
{\tilde{y}}^{\left(1\right)}\left(\Omega\right) & = & -\frac{i\sqrt{2\mu}\;\xi_{y1}\left(\Omega\right)}{\left(-i\Omega+1+\mu\right)}\nonumber \\
{\tilde{x}}^{+\left(1\right)}\left(\Omega\right) & = & \frac{\sqrt{2\mu}\;\xi_{x2}\left(\Omega\right)}{\left(-i\Omega+1-\mu\right)}\nonumber \\
{\tilde{y}}^{+\left(1\right)}\left(\Omega\right) & = & -\frac{i\sqrt{2\mu}\;\xi_{y2}\left(\Omega\right)}{\left(-i\Omega+1+\mu\right)}\end{eqnarray}

\item Second order: \begin{eqnarray}
{\tilde{x}}_{0}^{\left(2\right)}\left(\Omega\right) & = & -\frac{\gamma_{r}\left[\tilde{x}^{\left(1\right)}\star\tilde{x}^{+\left(1\right)}-\tilde{y}^{\left(1\right)}\star\tilde{y}^{+\left(1\right)}\right]\left(\Omega\right)}{\left(-i\Omega+\gamma_{r}\right)}\nonumber \\
{\tilde{y}}_{0}^{\left(2\right)}\left(\Omega\right) & = & -\frac{\gamma_{r}\left[\tilde{x}^{\left(1\right)}\star\tilde{y}^{+\left(1\right)}+\tilde{x}^{+\left(1\right)}\star\tilde{y}^{\left(1\right)}\right]\left(\Omega\right)}{\left(-i\Omega+\gamma_{r}\right)}\end{eqnarray}

\item Third order: 
\end{itemize}
\begin{eqnarray}
{\tilde{x}}^{\left(3\right)}\left(\Omega\right) & = & \frac{\left[\tilde{x}_{0}^{\left(2\right)}\star\left(\tilde{x}^{\left(1\right)}+\xi_{x1}/\sqrt{2\mu}\right)+\tilde{y}_{0}^{\left(2\right)}\star\left(\tilde{y}^{\left(1\right)}+i\xi_{y1}/\sqrt{2\mu}\right)\right]\left(\Omega\right)}{2\left(-i\Omega+1-\mu\right)}\nonumber \\
{\tilde{y}}^{\left(3\right)}\left(\Omega\right) & = & \frac{\left[\tilde{y}_{0}^{\left(2\right)}\star\left(\tilde{x}^{\left(1\right)}+\xi_{x1}/\sqrt{2\mu}\right)-\tilde{x}_{0}^{\left(2\right)}\star\left(\tilde{y}^{\left(1\right)}+i\xi_{y1}/\sqrt{2\mu}\right)\right]\left(\Omega\right)}{2\left(-i\Omega+1+\mu\right)}\nonumber \\
{\tilde{x}}^{+\left(3\right)}\left(\Omega\right) & = & \frac{\left[\tilde{x}_{0}^{\left(2\right)}\star\left(\tilde{x}^{+\left(1\right)}+\xi_{x2}/\sqrt{2\mu}\right)+\tilde{y}_{0}^{\left(2\right)}\star\left(\tilde{y}^{+\left(1\right)}+i\xi_{y2}/\sqrt{2\mu}\right)\right]\left(\Omega\right)}{2\left(-i\Omega+1-\mu\right)}\nonumber \\
{\tilde{y}}^{+\left(3\right)}\left(\Omega\right) & = & \frac{\left[\tilde{y}_{0}^{\left(2\right)}\star\left(\tilde{x}^{+\left(1\right)}+\xi_{x2}/\sqrt{2\mu}\right)-\tilde{x}_{0}^{\left(2\right)}\star\left(\tilde{y}^{+\left(1\right)}+i\xi_{y2}/\sqrt{2\mu}\right)\right]\left(\Omega\right)}{2\left(-i\Omega+1+\mu\right)}\end{eqnarray}

\subsection{Squeezing correlation spectrum}

We now calculate the spectrum of the squeezed field, which is given by $\left\langle \tilde{y}\left(\Omega_{1}\right)\tilde{y}^{+}\left(\Omega_{2}\right)\right\rangle $.

\begin{eqnarray}
\left\langle \tilde{y}\left(\Omega_{1}\right)\tilde{y}^{+}\left(\Omega_{2}\right)\right\rangle  & = & g^{2}\left\langle \tilde{y}^{\left(1\right)}\left(\Omega_{1}\right)\tilde{y}^{+\left(1\right)}\left(\Omega_{2}\right)\right\rangle +\nonumber \\
 &  & +g^{4}\left[\left\langle \tilde{y}^{\left(1\right)}\left(\Omega_{1}\right)\tilde{y}^{+\left(3\right)}\left(\Omega_{2}\right)\right\rangle +\left\langle \tilde{y}^{\left(3\right)}\left(\Omega_{1}\right)\tilde{y}^{+\left(1\right)}\left(\Omega_{2}\right)\right\rangle \right]+\cdots\end{eqnarray}

The lowest order contribution is the usual result of the linearized theory and given is given by: \begin{equation}
\left\langle \tilde{y}^{\left(1\right)}\left(\Omega_{1}\right)\tilde{y}^{+\left(1\right)}\left(\Omega_{2}\right)\right\rangle =-\frac{2\mu\delta\left(\Omega_{1}+\Omega_{2}\right)}{\left[\Omega_{1}^{2}+\left(1+\mu\right)^{2}\right]}\,\,.\end{equation}

In terms of the squeezing variance, this means that:

\begin{equation}
V^{(1)\pi/2}(\Omega)=1-\frac{4\mu}{\Omega^{2}+(1+\mu)^{2}}\end{equation}
For comparison, note that the complementary (unsqueezed) spectrum to this order is: \begin{equation}
\left\langle \widetilde{x}^{\left(1\right)}\left(\Omega_{1}\right)\widetilde{x}^{+\left(1\right)}\left(\Omega_{2}\right)\right\rangle =\frac{2\mu\delta\left(\Omega_{1}+\Omega_{2}\right)}{\left[\Omega_{1}^{2}+\left(1-\mu\right)^{2}\right]}\,\,.\end{equation}
 Taking the next order corrections into account we find that the spectrum of the squeezed quadrature is given by \begin{equation}
\left\langle \tilde{y}\left(\Omega_{1}\right)\tilde{y}^{+}\left(\Omega_{2}\right)\right\rangle =g^{2}\delta(\Omega_{1}+\Omega_{2})S(\Omega_{1})\,\,,\end{equation}
 where $S(\Omega)$ is given by: \begin{eqnarray}
S(\Omega) & = & \frac{-2\mu}{\Omega^{2}+(1+\mu)^{2}}+\frac{2g^{2}\mu^{2}\gamma_{r}}{[\Omega^{2}+(1+\mu)^{2}]^{2}}\times\nonumber \\
 &  & \times\left[\frac{\left(\Omega^{2}+1-\mu^{2}\right)}{\mu\gamma_{r}(1-\mu^{2})}+\frac{(1-\mu+\gamma_{r})(1+\mu)-\Omega^{2}}{(1-\mu)[\Omega^{2}+(1-\mu+\gamma_{r})^{2}]}\right.-\nonumber \\
 &  & \left.-\frac{(1+\mu+\gamma_{r})(1+\mu)-\Omega^{2}}{(1+\mu)[\Omega^{2}+(1+\mu+\gamma_{r})^{2}]}\right]\,\,.\end{eqnarray}
 The correctness of the above expression can be checked by verifying the following equality: \begin{eqnarray}
\left\langle {y}^{\left(1\right)}\left(\tau\right){y}^{+\left(3\right)}\left(\tau\right)\right\rangle _{ss}=\int\frac{d\Omega_{1}}{\sqrt{2\pi}}\int\frac{d\Omega_{2}}{\sqrt{2\pi}}~e^{i(\Omega_{1}+\Omega_{2})\tau}\left\langle \tilde{y}^{\left(1\right)}\left(\Omega_{1}\right)\tilde{y}^{+\left(3\right)}\left(\Omega_{2}\right)\right\rangle \,\,.\end{eqnarray}
 The corresponding external squeezing spectrum is then: \begin{eqnarray}
V^{\pi/2}(\Omega) & = & 1-\frac{4\mu}{\Omega^{2}+(1+\mu)^{2}}+\frac{4g^{2}\mu^{2}\gamma_{r}}{[\Omega^{2}+(1+\mu)^{2}]^{2}}\nonumber \\
 &  & \times\left[\frac{(\Omega^{2}+1-\mu^{2})}{\mu\gamma_{r}(1-\mu^{2})}+\frac{(1-\mu+\gamma_{r})(1+\mu)-\Omega^{2}}{(1-\mu)[\Omega^{2}+(1-\mu+\gamma_{r})^{2}]}\right.\nonumber \\
 &  & \left.-\frac{(1+\mu+\gamma_{r})(1+\mu)-\Omega^{2}}{(1+\mu)[\Omega^{2}+(1+\mu+\gamma_{r})^{2}]}\right]\,\,.\end{eqnarray}

This equation gives the complete squeezing spectrum, including all nonlinear correction to order $g^{2}$ or $1/N$. The
linear part gives perfect squeezing for $\mu=1$, and $\Omega=0$, as expected from the linear theory. The nonlinear terms
give corrections to perfect squeezing below threshold. At zero frequency, we find that:\begin{eqnarray}
V^{\pi/2}(0) & = & 1-\frac{4\mu}{(1+\mu)^{2}}+\nonumber \\
 & + & \frac{4g^{2}\mu}{(1+\mu)^{4}}\left[1+\frac{2\mu^{2}\gamma_{r}(2+\gamma_{r})}{(1-\mu)((1+\gamma_{r})^{2}-\mu^{2})}\right]\nonumber \\
\end{eqnarray}

\begin{figure}
\includegraphics[%
  width=8cm]{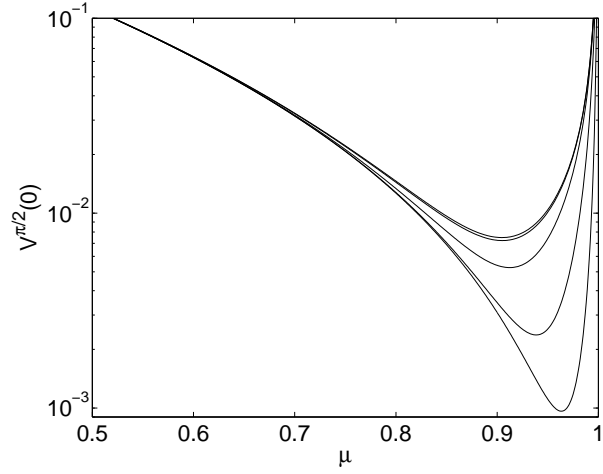}

\caption{Optimum squeezing with $g^{2}=0.001$, $\gamma_{r}=10^{-3},10^{-2},10^{-1},1,10$ . Upper lines have larger values of $\gamma_{r}$.
Here, $V^{\pi/2}<1$ indicates squeezing and entanglement occurring at zero frequency. \label{cap:Optimum-squeezing}}
\end{figure}

The resulting behavior for the optimum entanglement, which is found at zero-frequency (ignoring complications from technical
noise), is shown in Fig (\ref{cap:Optimum-squeezing}). We see that, as expected, the entanglement is not optimized at the
critical point, since the nonlinear critical fluctuations spoil this before an ideal entangled two-mode squeezed state with
$V^{\pi/2}=0$ is achieved. Better entanglement is obtained when $\gamma_{r}$ is reduced, as this minimizes the `information
leakage' in the losses of the pump mode. In this limit, the only losses are through the signal and idler output ports, which
are needed in order to have extra-cavity measurements.

This expression does not describe the spectrum very close to the critical point, as it diverges at the threshold. This region
requires a different kind of scaling and is discussed later.

The complementary or unsqueezed spectrum, for measurements of the maximum quadrature fluctuations, is given by:

\begin{eqnarray}
V^{0}(\Omega) & = & 1+\frac{4\mu}{\Omega^{2}+(1-\mu)^{2}}-\frac{4g^{2}\mu^{2}\gamma_{r}}{[\Omega^{2}+(1-\mu)^{2}]^{2}}\nonumber \\
 &  & \times\left[\frac{(\Omega^{2}+1-\mu^{2})}{\mu\gamma_{r}(1-\mu^{2})}+\frac{(1-\mu+\gamma_{r})(1-\mu)-\Omega^{2}}{(1-\mu)[\Omega^{2}+(1-\mu+\gamma_{r})^{2}]}\right.\nonumber \\
 &  & \left.-\frac{(1+\mu+\gamma_{r})(1-\mu)-\Omega^{2}}{(1+\mu)[\Omega^{2}+(1+\mu+\gamma_{r})^{2}]}\right]\,\,.\end{eqnarray}

\begin{figure}
\includegraphics[%
  width=8cm]{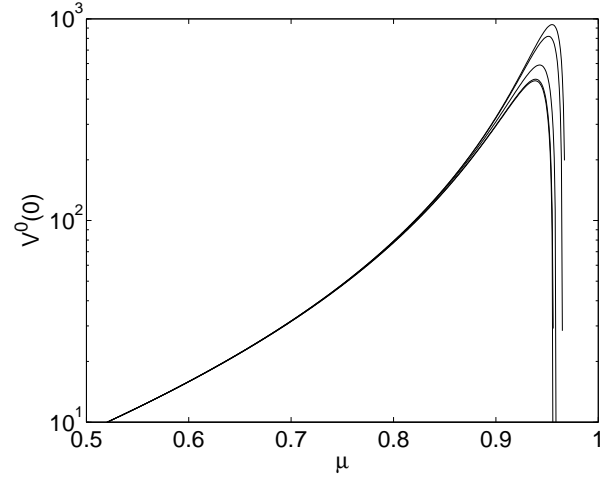}

\caption{Complementary (unsqueezed) spectrum with $g^{2}=0.001$, $\gamma_{r}=10^{-3},10^{-2},10^{-1},1,10$ . Lower lines have larger
values of $\gamma_{r}$.\label{cap:Optimum-unsqueezing}}
\end{figure}

The resulting behavior for the zero-frequency critical fluctuations is shown in Fig (\ref{cap:Optimum-unsqueezing}). Near
the critical point, higher order terms are likely to become significant. The effects of these are treated in the next section.

We note here that in the linearized analysis, the product of these spectra corresponds to the Heisenberg uncertainty principal:\begin{eqnarray*}
V^{0}(\Omega)V^{\pi/2}(\Omega) & = & \left[1-\frac{4\mu}{\Omega^{2}+(1+\mu)^{2}}\right]\left[1+\frac{4\mu}{\Omega^{2}+(1-\mu)^{2}}\right]\\
 & = & 1\,\,\,\,.\end{eqnarray*}
 Near threshold where nonlinear effects are dominant, this relationship no longer holds.%
\begin{figure}
\includegraphics[%
  width=8cm]{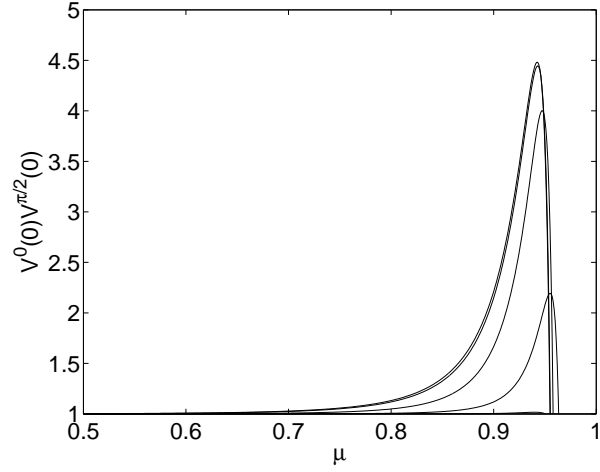}

\caption{Heisenberg uncertainty product with $g^{2}=0.001$, $\gamma_{r}=10^{-3},10^{-2},10^{-1},1,10$ . Upper lines have larger
values of $\gamma_{r}$.\label{Heisenberg}}
\end{figure}
The zero-frequency nonlinear uncertainty product is shown in Fig (\ref{Heisenberg}). Just below the critical point, the
nonlinear corrections apparently predict an uncertainty product less than unity, which is clearly the point at which the
second-order perturbation method breaks down. An unexpected feature of these results is that for $\gamma_{r}\ll1$ , the
uncertainty product remains close to unity for all driving fields, indicating that there is a near minimum uncertainty state
for low-frequency spectral measurements in the output fields. This does not mean that there is a minimum uncertainty state
for the internal quadrature moments, since these are effectively integrated over all frequencies, and involve different quantum
fields.

\begin{figure}
\includegraphics[%
  width=8cm]{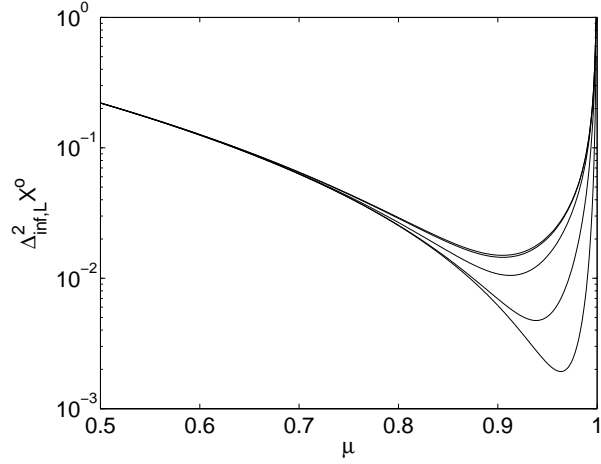}

\caption{Inferred quadrature uncertainty with $g^{2}=0.001$, $\gamma_{r}=10^{-3},10^{-2},10^{-1},1,10$ . Upper lines have larger
values of $\gamma_{r}$. When $\Delta_{inf,L}^{2}X^{o}<1$, one may infer an EPR paradox. \label{Inference}}
\end{figure}

We also investigate the behavior of the inferred Heisenberg uncertainty product, which demonstrates that there is an EPR
paradox. In the original proposal, this uncertainty product would be zero, as the original EPR paradox involved perfect correlations.
Instead, the minimum value of this product is determined by the nonlinear critical fluctuations. Due to symmetry, we only
need plot the behavior of $\Delta_{inf,L}^{2}X^{o}$ in Fig (\ref{Inference}).

This shows qualitatively similar behavior to the entanglement measure based on squeezing, and in fact for strong entanglement
the inferred uncertainty and squeezing measures are simply related by \[
\Delta_{inf,L}^{2}X^{o}=2V^{\pi/2}\,\,.\]

We see that near threshold, the EPR measure and squeezing entanglement measure both show the existence of a strongly entangled
output beam, as one might expect - but the perturbation theory breaks down past the point where optimum entanglement is achieved,
just below threshold.

\subsection{Triple Spectral Correlations}

Triple spectral correlations give quantum effects which distinguish very strongly\cite{Triple} between the full quantum
theory and the semi-classical approximation.

Here, we calculate the internal quadrature triple spectral correlation $\left\langle \tilde{x}\left(\Omega_{1}\right)\tilde{y}^{+}\left(\Omega_{2}\right)\tilde{y}_{0}\left(\Omega_{3}\right)\right\rangle $.
To the lowest non-vanishing order this is given by \begin{eqnarray}
\left\langle \tilde{x}\left(\Omega_{1}\right)\tilde{y}^{+}\left(\Omega_{2}\right)\tilde{y}_{0}\left(\Omega_{3}\right)\right\rangle  & = & g^{4}\left\langle \tilde{x}^{\left(1\right)}\left(\Omega_{1}\right)\tilde{y}^{+\left(1\right)}\left(\Omega_{2}\right)\tilde{y}_{0}^{\left(2\right)}\left(\Omega_{3}\right)\right\rangle \,\,.\end{eqnarray}

Substituting for $\tilde{y}_{0}^{\left(2\right)}$ , we have \begin{eqnarray}
\left\langle \tilde{x}^{\left(1\right)}\left(\Omega_{1}\right)\tilde{y}^{+\left(1\right)}\left(\Omega_{2}\right)\tilde{y}_{0}^{\left(2\right)}\left(\Omega_{3}\right)\right\rangle  & = & -\frac{\gamma_{r}\left\langle \tilde{x}^{\left(1\right)}\left(\Omega_{1}\right)\tilde{y}^{+\left(1\right)}\left(\Omega_{2}\right)[\tilde{x}^{\left(1\right)}\star\tilde{y}^{+\left(1\right)}+\tilde{x}^{+\left(1\right)}\star\tilde{y}^{\left(1\right)}](\Omega_{3})\right\rangle }{\left(-i\Omega_{3}+\gamma_{r}\right)}\,\,.\end{eqnarray}

and using the Gaussian nature of the stochastic variables involved to factorize the fourth order correlations we obtain:
\begin{equation}
\left\langle \tilde{x}^{\left(1\right)}\left(\Omega_{1}\right)\tilde{y}^{+\left(1\right)}\left(\Omega_{2}\right)\tilde{y}_{0}^{\left(2\right)}\left(\Omega_{3}\right)\right\rangle =\frac{4\mu^{2}\gamma_{r}/\sqrt{2\pi}\;\delta\left(\Omega_{1}+\Omega_{2}+\Omega_{3}\right)}{\left(-i\Omega_{3}+\gamma_{r}\right)\left[\Omega_{1}^{2}+\left(1-\mu\right)^{2}\right]\left[\Omega_{2}^{2}+\left(1+\mu\right)^{2}\right]}\,\,.\end{equation}

To check this result, we evaluate the steady state moment $\left\langle {x}^{\left(1\right)}\left(\tau\right){y}^{+\left(1\right)}\left(\tau\right){y}_{0}^{\left(2\right)}\left(\tau\right)\right\rangle _{ss}$
using \begin{eqnarray}
\left\langle {x}^{\left(1\right)}\left(\tau\right){y}^{+\left(1\right)}\left(\tau\right){y}_{0}^{\left(2\right)}\left(\tau\right)\right\rangle _{ss} & = & \int\frac{d\Omega_{1}}{}{\sqrt{2\pi}}\int\frac{d\Omega_{2}}{\sqrt{2\pi}}\int\frac{d\Omega_{3}}{\sqrt{2\pi}}~e^{i(\Omega_{1}+\Omega_{2}+\Omega_{3})\tau}\times\\
 &  & \times\left\langle \tilde{x}^{\left(1\right)}\left(\Omega_{1}\right)\tilde{y}^{+\left(1\right)}\left(\Omega_{2}\right)\tilde{y}_{0}^{\left(2\right)}\left(\Omega_{3}\right)\right\rangle \,\,.\end{eqnarray}
 and find that we obtain the same result as given earlier by direct calculations.

This result will be compared later with the corresponding result obtained in the semi-classical theory .

\subsection{Comparisons with simulations}

In order to verify the accuracy of these analytic calculations, we performed extensive numerical simulations of the full
nonlinear stochastic simulations, using a differencing technique as in earlier studies. We only calculate the nonlinear squeezing
variance, defined as:

\begin{equation}
V(\Omega)=V^{\pi/2}(\Omega)-V^{(1)\pi/2}(\Omega)\,\,.\end{equation}
This allows us to focus on the nonlinear corrections, which are relatively small except very near the critical threshold
at $\mu=1$. The numerical method has the advantage that, unlike perturbation theory, it is valid at all driving fields -
even at the critical point.

The integration parameters used were step size $d\tau=0.001$, with a time-window of $\tau_{max}=10000$. The number of stochastic
trajectories used for averaging were $2000$, resulting in typical relative sampling errors of around $\pm2\%$, as can be
seen from the background sampling noise in some of the resulting spectra. 

Typical results are shown in Figs (\ref{mu0.5}-\ref{mu0.9}) below, for driving fields of $\mu=0.5,0.9$. Note that these
graphs only include the nonlinear corrections. Excellent agreement is found with the analytically predicted results for these
values of driving field.

\begin{figure}
\includegraphics[%
  width=8cm]{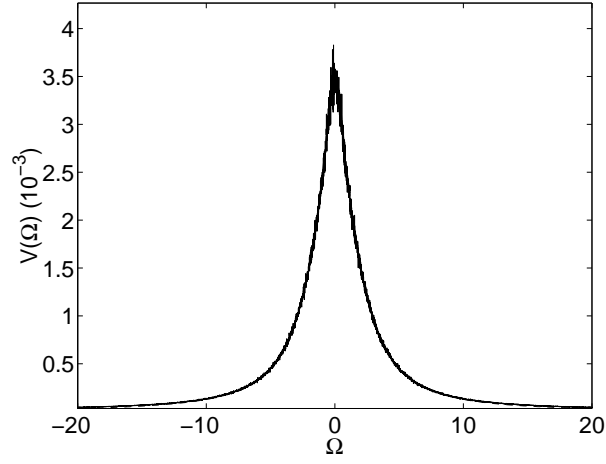}

\caption{Nonlinear squeezing spectrum with $g^{2}=0.005$, $\gamma_{r}=1$ and $\mu=0.5$. The dashed line represents the analytical
result and the noisy line the stochastic simulation.\label{mu0.5}}
\end{figure}

\begin{figure}
\includegraphics[%
  width=8cm]{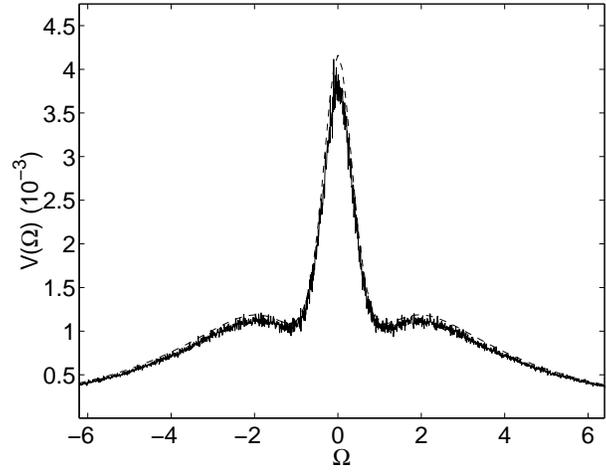}

\caption{Nonlinear squeezing spectrum with $g^{2}=0.001$, $\gamma_{r}=0.5$ and $\mu=0.9$. The dashed line represents the analytical
result and the noisy line the stochastic simulation.\label{mu0.9}}
\end{figure}

\begin{figure}
\includegraphics[%
  width=8cm]{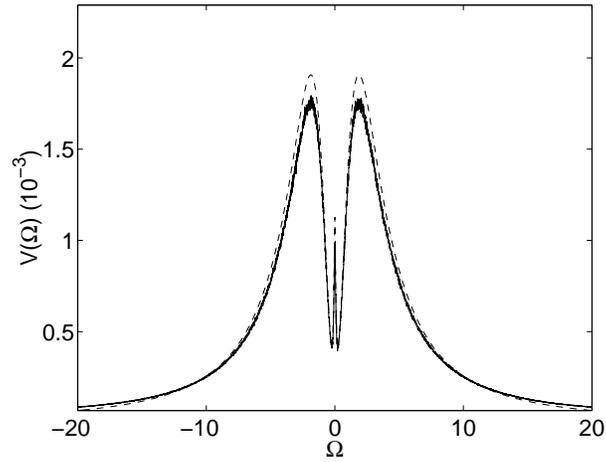}

\caption{Nonlinear squeezing spectrum with $g^{2}=0.001$, $\gamma_{r}=0.01$ and $\mu=0.93$. The dashed line represents the analytical
result and the noisy line the stochastic simulation. This is the driving field for optimum entanglement at zero frequency.\label{mu0.93}}
\end{figure}

Figure (\ref{mu0.93}) shows results slightly closer to threshold, at $\mu=0.93$, which is the optimum driving field for
the parameters chosen. 

At this point, a maximum error of around $10^{-4}$ is found, due to higher order nonlinear corrections. This indicates that
the analytic perturbation theory is able to correctly predict nonlinear effects up to the optimum squeezing point, but starts
to diverge beyond this point. The numerical results, however, are stable throughout the critical region. To obtain analytic
predictions in the critical region, we turn to a different asymptotic expansion in a later section.

\section{Semi-classical Spectral theory}

In this section we calculate approximate nonlinear results using a semi-classical approach. These are less reliable, especially
well below threshold, but have an intuitive `classical' interpretation in terms of the incoming vacuum fluctuations.

\subsection{Wigner Representation}

In the semi-classical theory, the hierarchy of the stochastic equations given earlier can be written, in the frequency domain,
as follows:

\begin{itemize}
\item First order \begin{eqnarray}
\tilde{x}_{0}^{(1)}(\Omega) & = & \frac{2\gamma_{r}\xi_{x0}(\Omega)}{\left(-i\Omega+\gamma_{r}\right)}\nonumber \\
\tilde{y}_{0}^{(1)}(\Omega) & = & \frac{2\gamma_{r}\xi_{y0}(\Omega)}{\left(-i\Omega+\gamma_{r}\right)}\nonumber \\
\tilde{x}^{(1)}(\Omega) & = & \frac{\sqrt{2}\xi_{x1}(\Omega)}{\left(-i\Omega+1-\mu\right)}\nonumber \\
\tilde{y}^{(1)}(\Omega) & = & \frac{\sqrt{2}\xi_{y1}(\Omega)}{\left(-i\Omega+1+\mu\right)}\nonumber \\
\tilde{x}^{+(1)}(\Omega) & = & \frac{\sqrt{2}\xi_{x2}(\Omega)}{\left(-i\Omega+1-\mu\right)}\nonumber \\
\tilde{y}^{+(1)}(\Omega) & = & \frac{\sqrt{2}\xi_{y2}(\Omega)}{\left(-i\Omega+1+\mu\right)}\end{eqnarray}

\item Second order \begin{eqnarray}
\tilde{x}_{0}^{(2)}(\Omega) & = & -\frac{\gamma_{r}\left[\tilde{x}^{(1)}\star\tilde{x}^{+(1)}-\tilde{y}^{(1)}\star\tilde{y}^{+(1)}\right](\Omega)}{\left(-i\Omega+\gamma_{r}\right)}\nonumber \\
\tilde{y}_{0}^{(2)}(\Omega) & = & -\frac{\gamma_{r}\left[\tilde{x}^{(1)}\star\tilde{y}^{+(1)}+\tilde{x}^{+(1)}\star\tilde{y}^{(1)}\right](\Omega)}{\left(-i\Omega+\gamma_{r}\right)}\nonumber \\
\tilde{x}^{(2)}(\Omega) & = & \frac{\left[\tilde{x}^{(1)}\star\tilde{x}_{0}^{(1)}+\tilde{y}^{(1)}\star\tilde{y}_{0}^{(1)}\right](\Omega)}{2\left(-i\Omega+1-\mu\right)}\nonumber \\
\tilde{y}^{(2)}(\Omega) & = & \frac{\left[\tilde{x}^{(1)}\star\tilde{y}_{0}^{(1)}-\tilde{y}^{(1)}\star\tilde{x}_{0}^{(1)}\right](\Omega)}{2\left(-i\Omega+1+\mu\right)}\nonumber \\
\tilde{x}^{+(2)}(\Omega) & = & \frac{\left[\tilde{x}^{+(1)}\star\tilde{x}_{0}^{(1)}+\tilde{y}^{+(1)}\star\tilde{y}_{0}^{(1)}\right](\Omega)}{2\left(-i\Omega+1-\mu\right)}\nonumber \\
\tilde{y}^{+(2)}(\Omega) & = & \frac{\left[\tilde{x}^{+(1)}\star\tilde{y}_{0}^{(1)}-\tilde{y}^{+(1)}\star\tilde{x}_{0}^{(1)}\right](\Omega)}{2\left(-i\Omega+1+\mu\right)}\end{eqnarray}

\item Third order (signal and idler fields) 
\end{itemize}
\begin{eqnarray}
\tilde{x}^{(3)}(\Omega) & = & \frac{\left[\tilde{x}^{(1)}\star\tilde{x}_{0}^{(2)}+\tilde{x}^{(2)}\star\tilde{x}_{0}^{(1)}+\tilde{y}^{(1)}\star\tilde{y}_{0}^{(2)}+\tilde{y}^{(2)}\star\tilde{y}_{0}^{(1)}\right](\Omega)}{2\left[-i\Omega+1-\mu\right]}\nonumber \\
\tilde{y}^{(3)}(\Omega) & = & \frac{\left[\tilde{x}^{(1)}\star\tilde{y}_{0}^{(2)}+\tilde{x}^{(2)}\star\tilde{y}_{0}^{(1)}-\tilde{y}^{(2)}\star\tilde{x}_{0}^{(1)}-\tilde{y}^{(1)}\star\tilde{x}_{0}^{(2)}\right](\Omega)}{2\left[-i\Omega+1+\mu\right]}\nonumber \\
\tilde{x}^{+(3)}(\Omega) & = & \frac{\left[\tilde{x}^{+(1)}\star\tilde{x}_{0}^{(2)}+\tilde{x}^{+(2)}\star\tilde{x}_{0}^{(1)}+\tilde{y}^{+(1)}\star\tilde{y}_{0}^{(2)}+\tilde{y}^{+(2)}\star\tilde{y}_{0}^{(1)}\right](\Omega)}{2\left[-i\Omega+1-\mu\right]}\nonumber \\
\tilde{y}^{+(3)}(\Omega) & = & \frac{\left[\tilde{x}^{+(1)}\star\tilde{y}_{0}^{(2)}+\tilde{x}^{+(2)}\star\tilde{y}_{0}^{(1)}-\tilde{y}^{+(2)}\star\tilde{x}_{0}^{(1)}-\tilde{y}^{+(1)}\star\tilde{x}_{0}^{(2)}\right](\Omega)}{2\left[-i\Omega+1+\mu\right]}\end{eqnarray}

\subsection{Squeezing Correlation spectrum}

The spectrum of the squeezed quadrature, for instance, is given by \begin{eqnarray}
\langle\tilde{y}(\Omega_{1})\tilde{y}^{+}(\Omega_{2})\rangle & = & g^{2}\langle\tilde{y}^{(1)}(\Omega_{1})\tilde{y}^{+(1)}(\Omega_{2})\rangle+g^{4}\left\{ \langle\tilde{y}^{(2)}(\Omega_{1})\tilde{y}^{+(2)}(\Omega_{2})\rangle\right.\nonumber \\
 & + & \left.\langle\tilde{y}^{(1)}(\Omega_{1})\tilde{y}^{+(3)}(\Omega_{2})\rangle+\langle\tilde{y}^{(3)}(\Omega_{1})\tilde{y}^{+(1)}(\Omega_{2})\rangle\right\} +\cdots\end{eqnarray}
 The lowest order contribution turns out to be \begin{equation}
\langle\tilde{y}^{(1)}(\Omega_{1})\tilde{y}^{+(1)}(\Omega_{2})\rangle=\frac{2\delta(\Omega_{1}+\Omega_{2})}{\Omega_{1}^{2}+(1+\mu)^{2}}\,\,,\end{equation}
 Similarly, for the amplified quadrature, to the lowest order, we have \begin{equation}
\langle\tilde{x}^{(1)}(\Omega_{1})\tilde{x}^{+(1)}(\Omega_{2})\rangle=\frac{2\delta(\Omega_{1}+\Omega_{2})}{\Omega_{1}^{2}+(1-\mu)^{2}}\,\,,\end{equation}
 For the pump quadratures, there is no squeezing, to the lowest order: \begin{equation}
\langle\tilde{x}_{0}^{(1)}(\Omega_{1})\tilde{x}_{0}^{(1)}(\Omega_{2})\rangle=\langle\tilde{y}_{0}^{(1)}(\Omega_{1})\tilde{y}_{0}^{(1)}(\Omega_{2})\rangle=\frac{4\gamma_{r}^{2}\delta(\Omega_{1}+\Omega_{2})}{\Omega_{1}^{2}+\gamma_{r}^{2}}\,\,.\end{equation}

The next contribution to the squeezed quadrature are \begin{eqnarray}
\langle\tilde{y}^{(2)}(\Omega_{1})\tilde{y}^{+(2)}(\Omega_{2})\rangle & = & \frac{\gamma_{r}\delta(\Omega_{1}+\Omega_{2})}{\Omega_{1}^{2}+(1+\mu)^{2}}\left\{ \frac{1-\mu+\gamma_{r}}{(1-\mu)\left[\Omega_{1}^{2}+(1-\mu+\gamma_{r})^{2}\right]}+\right.\nonumber \\
 &  & \left.\frac{1+\mu+\gamma_{r}}{(1+\mu)\left[\Omega_{1}^{2}+(1+\mu+\gamma_{r})^{2}\right]}\right\} \,\,,\end{eqnarray}

and \begin{eqnarray}
\langle\tilde{y}^{(1)}(\Omega_{1})\tilde{y}^{+(3)}(\Omega_{2})\rangle+\langle\tilde{y}^{(3)}(\Omega_{1})\tilde{y}^{+(1)}(\Omega_{2})\rangle & = & \frac{2\mu\gamma_{r}\delta(\Omega_{1}+\Omega_{2})}{\left[\Omega_{1}^{2}+(1+\mu)^{2}\right]^{2}}\times\nonumber \\
 &  & \left\{ -\frac{(1+\mu)(1-\mu+\gamma_{r})-\Omega_{1}^{2}}{(1-\mu)\left[\Omega_{1}^{2}+(1-\mu+\gamma_{r})^{2}\right]}\right.\nonumber \\
 &  & \left.+\frac{(1+\mu)(1+\mu+\gamma_{r})-\Omega_{1}^{2}}{(1+\mu)\left[\Omega_{1}^{2}+(1+\mu+\gamma_{r})^{2}\right]}+\frac{2(1+\mu)}{\gamma_{r}(1-\mu^{2})}\right\} \,\,,\end{eqnarray}

\noindent which yield, for the $S(\Omega)$\begin{eqnarray}
S(\Omega) & = & \frac{2}{\Omega^{2}+(1+\mu)^{2}}+\frac{g^{2}\gamma_{r}}{\left[\Omega^{2}+(1+\mu)^{2}\right]^{2}}\left\{ \frac{4\mu(1+\mu)}{\gamma_{r}(1-\mu^{2})}+\right.\nonumber \\
 &  & +\left.\frac{(1-\mu+\gamma_{r})\Omega^{2}+\left[(1+\mu)^{2}+2\mu(1+\mu)\right](1+\mu+\gamma_{r})}{(1+\mu)\left[\Omega^{2}+(1+\mu+\gamma_{r})^{2}\right]}\right.\nonumber \\
 &  & \left.+\frac{(1+\mu+\gamma_{r})\Omega^{2}+(1-\mu^{2})(1-\mu+\gamma_{r})}{(1-\mu)\left[\Omega^{2}+(1-\mu+\gamma_{r})^{2}\right]}\right\} \,\,.\end{eqnarray}

This, in turn, gives the following expression for the external squeezing spectrum, obtained by including both internal fields
and the correlated reflected vacuum noise : \begin{eqnarray}
V^{\pi/2}(\Omega) & = & 1-\frac{4\mu}{\Omega^{2}+(1+\mu)^{2}}+\frac{2g^{2}\gamma_{r}}{\left[\Omega^{2}+(1+\mu)^{2}\right]^{2}}\left\{ \frac{2\mu(1+\Omega^{2}-\mu^{2})}{\gamma_{r}(1-\mu^{2})}\right.\nonumber \\
 &  & +\left.\frac{\left[(1-\mu)(1-\mu+\gamma_{r})-2\mu^{2}\right]\Omega^{2}+(1-\mu+\gamma_{r})\left(1+\mu+\mu^{2}+\mu^{3}\right)}{(1-\mu)\left[\Omega^{2}+(1-\mu+\gamma_{r})^{2}\right]}\right.\nonumber \\
 &  & +\left.\frac{\left[(1+\mu)(1+\mu+\gamma_{r})+2\mu^{2}\right]\Omega^{2}+(1+\mu+\gamma_{r})\left(1+3\mu+\mu^{2}-\mu^{3}\right)}{(1+\mu)\left[\Omega^{2}+(1+\mu+\gamma_{r})^{2}\right]}\right.\,\,.\end{eqnarray}

It is interesting to note that this spectrum is quite different from that given by positive P-representation when $\mu\rightarrow0$.
However near the threshold, that is in the limit $\mu\rightarrow1$, the two results show close agreement. This means that
even when the pump is off, the semi-classical theory gives a distorted vacuum spectrum due to the presence of the nonlinear
crystal. This happens because in this theory the vacuum fluctuations are taken as real, and then two vacuum modes can interact
inside the crystal as real fields. In the limit of $\gamma_{r}\rightarrow0$, the two spectra again become compatible, as
the semi-classical theory decouples the fundamental mode from its vacuum input in this limit. In the case of threshold fluctuations,
we can interpret the agreement as due to the fact that in this region large numbers of photon numbers involved - which means
that the truncation approximation used in the semi-classical approximation is fairly reliable.

\subsection{Triple spectral correlation}

For the triple spectral correlation function \begin{eqnarray}
\langle\tilde{x}(\Omega_{1})\tilde{y}^{+}(\Omega_{2})\tilde{y}_{0}(\Omega_{3})\rangle & = & g^{3}\langle\tilde{x}^{(1)}(\Omega_{1})\tilde{y}^{+(1)}(\Omega_{2})\tilde{y}_{0}^{(1)}(\Omega_{3})\rangle+g^{4}\left\{ \langle\tilde{x}^{(1)}(\Omega_{1})\tilde{y}^{+(1)}(\Omega_{2})\tilde{y}_{0}^{(2)}(\Omega_{3})\rangle\right.\nonumber \\
 &  & \left.+\langle\tilde{x}^{(2)}(\Omega_{1})\tilde{y}^{+(1)}(\Omega_{2})\tilde{y}_{0}^{(1)}(\Omega_{3})\rangle+\langle\tilde{x}^{(1)}(\Omega_{1})\tilde{y}^{+(2)}(\Omega_{2})\tilde{y}_{0}^{(1)}(\Omega_{3})\rangle\right\} \,\,.\end{eqnarray}

the term proportional to $g^{3}$ vanishes, and the result, to the lowest non-trivial order is found to be

\begin{eqnarray}
\langle\tilde{x}(\Omega_{1})\tilde{y}^{+}(\Omega_{2})\tilde{y}_{0}(\Omega_{3})\rangle & = & g^{4}\frac{4\gamma_{r}\delta(\Omega_{1}+\Omega_{2}+\Omega_{3})}{\sqrt{2\pi}}\times\nonumber \\
 & \times & \left\{ -\frac{1}{(-i\Omega_{3}+\gamma_{r})[\Omega_{1}^{2}+(1-\mu)^{2}][\Omega_{2}^{2}+(1+\mu)^{2}]}+\right.\nonumber \\
 & + & \left.\frac{\gamma_{r}}{(-i\Omega_{1}+1-\mu)[\Omega_{2}^{2}+(1+\mu)^{2}][\Omega_{3}^{2}+\gamma_{r}^{2}]}+\right.\nonumber \\
 & + & \left.\frac{\gamma_{r}}{(-i\Omega_{2}+1+\mu)[\Omega_{1}^{2}+(1-\mu)^{2}][\Omega_{3}^{2}+\gamma_{r}^{2}]}\right\} \,\,.\end{eqnarray}

As before, the essential difference between quantum and semi-classical theories is that the former gives a zero spectrum
in the absence of a driving field while the latter, due to the real character of the vacuum field, gives a non zero correlation.

\section{Critical Perturbation Theory}

As we have seen, the perturbative corrections diverge at the critical point ($\mu=1$) and a new approach is called for to
investigate the neighborhood of the threshold. To this end we define new scaled quadratures variables, and use a different
expansion \cite{Critical} valid around the critical region. The new pump mode variable $x_{0}$ now corresponds to the real
scaled depletion in the pump mode amplitude, relative to the undepleted value at the critical point. The signal-idler quadrature
variables $x\;,\; x^{+}$ now describe the critical fluctuations scaled to be of order $1$ at the threshold.

\subsection{Positive-P Representation}

We scale the quadratures as \begin{eqnarray}
x_{0} & = & \frac{1}{g}\left[\frac{\chi X_{0}}{\gamma}-2\right]\,\,,\,\, y_{0}=\sqrt{\frac{2\gamma_{r}}{g}}Y_{0}\nonumber \\
x & = & \sqrt{g}X\,\,\,\,\,\,\,\,\,\,\,,\,\,\,\,\,\,\,\,\,\, y=Y\nonumber \\
x^{+} & = & \sqrt{g}X^{+}\,\,\,\,\,\,\,\,\,\,\,,\,\,\,\,\,\,\,\,\,\, y^{+}=Y^{+}\label{a35}\end{eqnarray}
 and define also a new scaled time and driving field \begin{eqnarray}
\eta & = & \frac{2}{g}\left(\frac{\mathcal{E}}{\mathcal{E}_{c}}-1\right)\nonumber \\
\tau & = & \gamma gt\label{b35}\end{eqnarray}

In terms of these variables, the equations in positive-P become \begin{eqnarray}
gdx_{0} & = & -\gamma_{r}\left[x_{0}-2\eta+xx^{+}-gyy^{+}\right]d\tau\nonumber \\
gdy_{0} & = & -\gamma_{r}\left[y_{0}+xy^{+}+yx^{+}\right]d\tau\nonumber \\
dx & = & \frac{1}{2}\left(x_{0}x+gy_{0}y\right)d\tau+dw_{x1}(\tau)\nonumber \\
gdy & = & \left[-2y+\frac{g}{2}\left(xy_{0}-yx_{0}\right)\right]d\tau+dw_{y1}(\tau)\nonumber \\
dx^{+} & = & \frac{1}{2}\left(x_{0}x^{+}+gy_{0}y^{+}\right)d\tau+dw_{x2}(\tau)\nonumber \\
gdy^{+} & = & \left[-2y^{+}+\frac{g}{2}\left(x^{+}y_{0}-y^{+}x_{0}\right)\right]d\tau+dw_{y2}(\tau)\label{a36}\end{eqnarray}
 The Gaussian white noise sources in these equations are no longer uncorrelated and have the follow the properties \begin{eqnarray}
\langle dw_{x1}dw_{x2}\rangle & = & 2\left(1+\frac{g}{2}x_{0}\right)d\tau\nonumber \\
\langle dw_{y1}dw_{y2}\rangle & = & -2g\left(1+\frac{g}{2}x_{0}\right)d\tau\nonumber \\
\langle dw_{x1}dw_{y2}\rangle & = & \langle dw_{x2}dw_{y1}\rangle=g^{2}y_{0}d\tau\label{a37}\end{eqnarray}
 We now develop a perturbation theory valid near the threshold. The first set of equations is obtained by neglecting all
terms of order $g$ or greater on the right sides of the two sets of equations given above \begin{eqnarray}
gdx_{0}^{(0)} & = & -\gamma_{r}\left[x_{0}^{(0)}-2\eta+x^{(0)}x^{+(0)}\right]d\tau\nonumber \\
gdy_{0}^{(0)} & = & -\gamma_{r}\left[y_{0}^{(0)}+x^{(0)}y^{+(0)}+y^{(0)}x^{+(0)}\right]d\tau\nonumber \\
dx^{(0)} & = & \frac{1}{2}\left[x^{(0)}x_{0}^{(0)}\right]d\tau+dw_{x1}^{(0)}\nonumber \\
gdy^{(0)} & = & -2y^{(0)}d\tau+dw_{y1}^{(0)}\nonumber \\
dx^{+(0)} & = & \frac{1}{2}\left[x^{+(0)}x_{0}^{(0)}\right]d\tau+dw_{x2}^{(0)}\nonumber \\
gdy^{+(0)} & = & -2y^{+(0)}d\tau+dw_{y2}^{(0)}\label{a38}\end{eqnarray}

A significant feature of these equations is that the quadratures $y^{(0)}$, and $y^{+(0)}$, can be worked out without reference
to any of other variables, and they give zero noise in the external quadrature at zero frequency. Coupling between variables
appears in high orders expansion and generates the critical fluctuations in the squeezed quadrature.

We now consider what happens at or near the classical threshold $\eta=0$. In a model where the sub-harmonic generation does
not cause the pump mode to deplete, we would have $x_{0}^{\left(0\right)}=2\eta$, and at threshold the critical fluctuations
in $x$ and $x^{+}$ would diffuse outward without any bound. When depletion is included, the critical fluctuations in these
quadratures are finite, but very slowly varying compared to those in the other variables. The pump field can therefore be
adiabatically eliminated to first order in the expansion.

Near threshold ($g\eta\ll1$) the decay term in the un-squeezed quadrature $x$ and $x^{+}$ is roughly $-x_{0}$, which
is of order $1$. The pump mode will be depleted, so $x_{0}$ must be negative in order for this to be stable. The scaled
pump field decay is $\gamma_{r}/g$, and the squeezed quadrature decay is of order $1/g$. If $\gamma_{r}$ is much larger
than $g$, it is possible to adiabatically eliminate both the pump amplitude and the squeezed quadrature in the equations
for the large critical fluctuations $x$ and $x^{+}$. Since we are taking the limit of small $g$, we shall assume that
this is possible to zero-th order in the asymptotic expansion. In the adiabatic elimination, we must solve for the steady
state values of the pump $x_{0}$, given an instantaneous first order critical fluctuation $x$ and $x^{+}$. To leading
(zeroth) order this gives \begin{equation}
x_{0}^{(0)}=2\eta-x^{(0)}x^{+(0)}\label{a39}\end{equation}

Substituting in the equations for $x$ and $x^{+}$, we find that \begin{eqnarray}
dx^{(0)} & = & \left[\eta x^{(0)}-\frac{1}{2}\left(x^{(0)}\right)^{2}x^{+(0)}\right]d\tau+dw_{x1}^{(0)}\nonumber \\
dx^{+(0)} & = & \left[\eta x^{+(0)}-\frac{1}{2}\left(x^{+(0)}\right)^{2}x^{(0)}\right]d\tau+dw_{x2}^{(0)}\label{a40}\end{eqnarray}

After the following change of variables \begin{equation}
x_{+}=\frac{x^{(0)}+x^{+(0)}}{2}\;\;\;\;\;\;\;\; x_{-}=i\frac{x^{(0)}-x^{+(0)}}{2}\label{a41}\end{equation}

the equations (\ref{a40}) can be put in the form \begin{equation}
\dot{\textbf{{x}}}=-\eta{\textbf{{x}}}-\frac{1}{2}{\textbf{{x}}}\left({\textbf{{x}}}\cdot{\textbf{{x}}}\right)+\xi(t)\label{a42}\end{equation}
 where ${\textbf{{x}}}$ is a two component vector whose elements are $x_{+}$ and $x_{-}$.

It is possible to write the Fokker-Planck equation for the probability density $P(x_{+},x_{-},t)$, and look for the equilibrium
distribution is of the form $P({\textbf{{x}}})=Nexp[-U({\textbf{{x}}})]$, where $U({\textbf{{x}}})$ is a potential function
given by \begin{equation}
U({\textbf{{x}}})=\eta{\textbf{{x}}}\cdot{\textbf{{x}}}+\frac{1}{4}\left({\textbf{{x}}}\cdot{\textbf{{x}}}\right)^{2}\label{a43}\end{equation}

The variance of the critical fluctuations at the critical point, $\eta=0$, is given by \begin{equation}
\langle x^{(0)}x^{+(0)}\rangle=2\frac{\Gamma(1)}{\Gamma(1/2)}=1.1284\label{a44}\end{equation}

\subsection{Critical Squeezing in positive-P Representation}

We can now find the steady state variance of the squeezed quadrature at threshold. Because the fluctuations in the squeezed
quadrature are very small, we must work to higher order in the asymptotic expansion to obtain a non trivial result. To achieve
this, it is most useful to introduce equations in the higher order moments $yy^{+}$ and $z=x^{+}y+xy^{+}$. The corresponding
stochastic equations are derived using It\^{o} rules for the variable changes, so that \begin{eqnarray}
gd(yy^{+}) & = & -2\left[1+2yy^{+}+\frac{g}{2}\left(x_{0}+x_{0}yy^{+}-\frac{1}{2}y_{0}z\right)\right]d\tau+ydw_{y2}+y^{+}dw_{y1}\nonumber \\
gdz & = & \left[-2z+\frac{g}{2}y_{0}\left(2xx^{+}+2gyy^{+}+4g\right)\right]d\tau+xdw_{y2}+x^{+}dw_{y1}+gydw_{x2}+gy^{+}dw_{x1}\label{a45}\end{eqnarray}

Taking the expectation value at the steady-state $\langle d(yy^{+})\rangle=0$, we get the first order correction \begin{equation}
\langle yy^{+}\rangle^{(1)}=-\frac{g}{4}\langle\left(1+yy^{+}\right)x_{0}-\frac{1}{2}y_{0}z\rangle^{(0)}\label{a46}\end{equation}

The first term in the above expression gives the result \begin{equation}
\langle\left(1+yy^{+}\right)x_{0}\rangle^{(0)}=\frac{1}{2}\langle x_{0}\rangle^{(0)}=\eta-\frac{1}{2}\langle x^{(0)}x^{+(0)}\rangle\label{a47}\end{equation}

For the second term we must write the correlation from the following equation \begin{equation}
gd(y_{0}z)=-\left[(2+\gamma_{r})y_{0}z+\gamma_{r}z^{2}\right]d\tau+0(g)+noise\label{a48}\end{equation}
 and then we get \begin{equation}
\langle y_{0}z\rangle^{(0)}=-\frac{\gamma_{r}}{2+\gamma_{r}}\langle z^{2}\rangle^{(0)}=-\frac{\gamma_{r}}{2+\gamma_{r}}\langle\left(x^{+}y+xy^{+}\right)^{2}\rangle^{(0)}=\frac{\gamma_{r}}{2+\gamma_{r}}\langle x^{(0)}x^{+(0)}\rangle\label{a49}\end{equation}
 So, finally we get, to first order, \begin{eqnarray}
\langle yy^{+}\rangle & = & \frac{1}{2}-\frac{g}{4}\left(\eta-\frac{1}{2}\langle x^{(0)}x^{+(0)}\rangle\right)+\frac{g}{8}\left(\frac{\gamma_{r}}{2+\gamma_{r}}\right)\langle x^{(0)}x^{+(0)}\rangle\nonumber \\
 & = & \frac{1}{2}-\frac{g\eta}{4}+\frac{g}{8}\left(\frac{2+2\gamma_{r}}{2+\gamma_{r}}\right)\langle x^{(0)}x^{+(0)}\rangle\label{a50}\end{eqnarray}

This result shows that the best squeezing, in the \emph{overall} moment, for the intra-cavity combined mode quadrature occurs
just above threshold, in much the same way as in the degenerate OPO \cite{CDD}.

\subsection{Wigner Representation}

As in the positive-P equations, we define new scaled quadratures variables to avoid divergences at the critical point \begin{eqnarray}
x_{0} & = & \frac{1}{g}\left[\frac{\chi X_{0}}{\gamma}-2\right]\,\,,\,\, y_{0}=\sqrt{2\gamma_{r}}Y_{0}\nonumber \\
x & = & \sqrt{g}X\,\,\,\,\,\,\,\,\,\,\,,\,\,\,\,\,\,\,\,\,\, y=Y\nonumber \\
x^{+} & = & \sqrt{g}X^{+}\,\,\,\,\,\,\,\,\,\,\,,\,\,\,\,\,\,\,\,\,\, y^{+}=Y^{+}\label{a51}\end{eqnarray}

In these new variables, the stochastic equations in the Wigner representation are ) \begin{eqnarray}
gdx_{0} & = & -\gamma_{r}\left[x_{0}-2\eta+xx^{+}-gyy^{+}\right]d\tau+dw_{x0}(\tau)\nonumber \\
gdy_{0} & = & -\gamma_{r}\left[y_{0}+\sqrt{g}\left(xy^{+}+yx^{+}\right)\right]d\tau+dw_{y0}(\tau)\nonumber \\
dx & = & \frac{1}{2}\left(x_{0}x+\sqrt{g}y_{0}y\right)d\tau+dw_{x1}(\tau)\nonumber \\
gdy & = & \left[-2y+\frac{1}{2}\left(\sqrt{g}xy_{0}-gyx_{0}\right)\right]d\tau+dw_{y1}(\tau)\nonumber \\
dx^{+} & = & \frac{1}{2}\left(x_{0}x^{+}+\sqrt{g}y_{0}y^{+}\right)d\tau+dw_{x2}(\tau)\nonumber \\
gdy^{+} & = & \left[-2y^{+}+\frac{1}{2}\left(\sqrt{g}x^{+}y_{0}-gy^{+}x_{0}\right)\right]d\tau+dw_{y2}(\tau)\label{a52}\end{eqnarray}
 Here we use the same notation for scaled time and driving field as in the positive-P case. The noise correlation are given
by \begin{eqnarray}
\langle dw_{x0}dw_{y0}\rangle & = & 4\gamma_{r}^{2}gd\tau\nonumber \\
\langle dw_{x1}dw_{x2}\rangle & = & 2d\tau\nonumber \\
\langle dw_{y1}dw_{y2}\rangle & = & 2gd\tau\label{a53}\end{eqnarray}

To develop a perturbation scheme, we define the zero order approximation to be the one in which terms of order and greater
than $\sqrt{g}$ are neglected in the set of equations above \begin{eqnarray}
gdx_{0}^{(0)} & = & -\gamma_{r}\left[x_{0}^{(0)}-2\eta+x^{(0)}x^{+(0)}\right]d\tau+dw_{x0}^{(0)}\nonumber \\
gdy_{0}^{(0)} & = & -\gamma_{r}\left[y_{0}^{(0)}+x^{(0)}y^{+(0)}+y^{(0)}x^{+(0)}\right]d\tau+dw_{y0}^{(0)}\nonumber \\
dx^{(0)} & = & \frac{1}{2}\left[x^{(0)}x_{0}^{(0)}\right]d\tau+dw_{x1}^{(0)}\nonumber \\
gdy^{(0)} & = & -2y^{(0)}d\tau+dw_{y1}^{(0)}\nonumber \\
dx^{+(0)} & = & \frac{1}{2}\left[x^{+(0)}x_{0}^{(0)}\right]d\tau+dw_{x2}^{(0)}\nonumber \\
gdy^{+(0)} & = & -2y^{+(0)}d\tau+dw_{y2}^{(0)}\label{a54}\end{eqnarray}

It is worth noting that this set of equations, though having the same structure as that in the positive-P case, has differences
in the correlations of the noise terms. On adiabatic elimination of the pump and substituting this result into $x^{0}$ and
$x^{+0}$ we find the same equations as in the positive-P representation, since to zero-th order the correlation noise in
both theories is identical.

\subsection{Critical squeezing in Wigner representation}

Now we proceed to calculate $\langle yy^{+}\rangle$ at threshold using the Wigner representation. Using the It\^{o} rules
we get \begin{equation}
gd(yy^{+})=2-4yy^{+}+\frac{\sqrt{g}}{2}y_{0}z-\frac{g}{2}2yy^{+}x_{0}+dw_{y1}+dw_{y2}\label{a55}\end{equation}

\noindent where we have defined $z=yx^{+}+y^{+}x$, which obey the following equation \begin{equation}
gdz=-2z+\sqrt{g}y_{0}xx^{+}+g\sqrt{g}y_{0}yy^{+}+x^{+}dw_{y1}+gydw_{x2}+xdw_{y2}+gy^{+}dw_{x1}\label{a56}\end{equation}

The squeezing variance at threshold in the steady state is obtained from the above equation taking expectation values \begin{equation}
\langle yy^{+}\rangle=\frac{1}{2}+\frac{\sqrt{g}}{8}\langle y_{0}z\rangle-\frac{g}{4}\langle x_{0}yy^{+}\rangle\label{a57}\end{equation}

The last term of the above equation can be written as \begin{equation}
\frac{g}{4}\langle x_{0}^{(0)}\rangle\langle yy^{+}\rangle^{(0)}=\frac{g\eta}{4}-\frac{g}{8}\langle x^{(0)}x^{+(0)}\rangle\label{a58}\end{equation}
 and the equation (55) gives the result \begin{equation}
\langle y_{0}z\rangle^{(0)}=-\sqrt{g}\frac{\gamma_{r}\langle z^{2}\rangle^{(0)}}{2+\gamma_{r}}+\sqrt{g}\frac{\langle y_{0}^{2}\rangle^{(0)}\langle x^{(0)}x^{+(0)}\rangle}{2+\gamma_{r}}\label{a59}\end{equation}

Using the results derived from the zero order equations \begin{eqnarray}
\langle y_{0}^{2}\rangle^{(0)} & = & 2\gamma_{r}\nonumber \\
\langle z^{2}\rangle^{(0)} & = & 2\langle x^{(0)}x^{+(0)}\rangle\langle y^{(0)}y^{+(0)}\rangle\label{a60}\end{eqnarray}
 we finally get \begin{equation}
\langle yy^{+}\rangle=\frac{1}{2}-\frac{g\eta}{4}+\frac{g}{8}\left(\frac{2+2\gamma_{r}}{2+\gamma_{r}}\right)\langle x^{(0)}x^{+(0)}\rangle\label{a61}\end{equation}

This result is exactly the same as obtained in positive P-representation. We can infer that dropping third order terms in
the Wigner phase space equation does not have any direct consequence for the near threshold analysis of entanglement to this
order of approximation. This is to be contrasted with the situation far below threshold, where there are large differences
in the nonlinear contributions, indicating a failure of the truncated (hidden-variable) Wigner theory. 

The change in behavior has a simple mathematical origin. Far below threshold, the signal/idler photon numbers are small,
which leads to a failure of the truncation approximation when using the semi-classical method. At the critical point, photon
numbers in all modes are relatively large, so the truncation approximation has less severe consequences.

\section{Conclusions}

We have calculated the effects of nonlinear quantum fluctuations in a nondegenerate parametric oscillator, both below and
at the classical threshold, using stochastic equations that follow from the positive P-representation. The analytical results
thus obtained are compared with exact numerical simulations. The spectral entanglement and squeezing in the output fields
is maximized just below threshold. This may be useful, for example, in cryptographic applications\cite{crycont}. We find
that at the critical point ($\mu=1$), the scaling behavior is quite different to the behavior below threshold, and must
be calculated by using an asymptotic perturbation theory, valid at the threshold itself. The total intra-cavity squeezing
and entanglement moment is actually minimized at a driving field just above threshold. This behavior was confirmed in our
simulations. This apparent paradox can be attributed to the fact that the critical fluctuations mostly tend to broaden the
squeezing spectrum, which has a strong effect at zero-frequency but does not diminish the total squeezing moment, integrated
over all frequencies.

A similar analysis was carried out within the framework of the semi-classical theory arising from a truncation to a Fokker-Planck
form of the evolution equation in the Wigner representation. Here, we found that well below threshold, while the linear terms
agreed with full quantum calculation, the nonlinear corrections tend to disagree, especially for low sub-harmonic losses.
However, at the critical point, the situation changes. Here, where the dominant terms are nonlinear, we find excellent agreement
between the two methods. While quantum fluctuations are indeed large at the critical point, it appears that an equally acceptable
interpretation of the observed noise characteristics near the critical point exists via a semi-classical model, which is
essentially a kind of hidden-variable theory.

Our main result is that entanglement, EPR correlations and squeezing are optimized very near threshold. At the same time,
the semi-classical Wigner approximation can give an excellent description of the squeezing and entanglement fluctuations
near threshold. On the other hand, some highly nonclassical signatures of quantum correlations occur in the higher-order
correlations, which are not described by the semi-classical approach. Surprisingly, these nonclassical and non-Gaussian signatures
only occur well below threshold, where one might have expected the usual linearized analysis to be applicable.

This suggests that experimental tests of the present theory may be carried out either near threshold - where the largest
effects will be observed in the enhanced critical fluctuations of the unsqueezed quadrature - or well below threshold, where
nonclassical triple correlations are predicted.

\vspace{1cm}

\begin{center}\textbf{ACKNOWLEDGMENT}\end{center}

\vspace{0.3cm}

We gratefully acknowledge financial support from CNPq (Brazil) and the Australian Research Council.


\begin{thebibliography}{10}
\bibitem{Yariv}A. Yariv, \textit{Quantum Electronics} (New York: Wiley), 1989. 
\bibitem{WKHW86}L. A. Wu, H. J. Kimble, J. L. Hall, H. Wu, Phys. Rev. Lett. \textbf{57}, 2520 (1986). 
\bibitem{HHRG87}A. Heidmann, R. J. Horowicz, S. Reynaud, E. Giacobino, C. Fabre, G. Camy, Phys. Rev. Lett. \textbf{59}, 2555 (1987). 
\bibitem{Hong}C. K. Hong, Z. Y. Ou, and L. Mandel, Phys. Rev. Lett. \textbf{59}, 2044 (1987).
\bibitem{Ou}Z. Y. Ou, S. F. Pereira, H. J. Kimble and K. C. Peng, Phys. Rev. Lett. \textbf{68}, 3663 (1992). 
\bibitem{yhang}Yun Zhang, Hai Wang, Xiaoying Li,Jietai Jing, Changde Xie and Kunchi Peng, Phys. Rev. A \textbf{62}, 023813 (2000). 
\bibitem{bow}W. P. Bowen, R. Schnabel, P. K. Lam, T. C. Ralph, Phys. Rev. Lett. 90 (4), 043601 (2003). 
\bibitem{silber}Ch. Silberhorn, P. K. Lam, O. Weiss, F. Konig, N. Korolkova and G. Leuchs, Phys. Rev. Lett. \textbf{86}, 4267 (2001). 
\bibitem{epr}A. Einstein, B. Podolsky and N. Rosen, Phys. Rev. \textbf{47}, 777, (1935). 
\bibitem{eprquad}M. D. Reid and P. D. Drummond, Phys. Rev. Lett. 60, 2731, (1988). P. Grangier, M. J. Potasek and B. Yurke, Phys. Rev. A \textbf{38},
3132, (1988). B. J. Oliver and C. R. Stroud, Phys. Lett. A \textbf{135}, 407, (1989). 
\bibitem{eprr}M. D. Reid, Phys. Rev. A \textbf{40}, 913 (1989).; ibid,quant-ph 0112038. 
\bibitem{rd}M. D. Reid and P.D. Drummond, Phys. Rev. A40, 4493 (1989), P. D. Drummond and M. D. Reid, Phys. Rev. A41, 3930 (1990). 
\bibitem{Pfister}S. Feng and O. Pfister, Journ. Opt. B, 5(3), 262 (2003); S. Feng and O. Pfister, quant-ph/0310002 (2003).
\bibitem{content}L. M. Duan, G. Giedke, J. I. Cirac and P. Zoller, Phys. Rev. Lett. \textbf{84}, 2722 (2000); R. Simon, Phys. Rev. Lett. \textbf{84},
2726 (2000). 
\bibitem{cventnat}N. Korolkova, C. Silberhorn, O. Glockl, et al. Eur. Phys. J D 18 (2): 229-235 (2002). 
\bibitem{cventpol}C. Schori, J. L. Sorensen, E. S. Polzik, Phys. Rev. A 66 (3) 033802 (2002). 
\bibitem{TE}T. W. Marshall, and E. Santos, Phys. Rev. A \textbf{41}, 1582 (1990). 
\bibitem{KTE}K. Dechoum, T.W. Marshall, and E. Santos, J. Mod. Opt., \textbf{47}, 1273 (2000). 
\bibitem{Olsen}M. K. Olsen, S. C. G. Granja, and R. J. Horowicz, Optics Comm. \textbf{165}, 293 (1999). 
\bibitem{Dechoum}M. K. Olsen, K. Dechoum and L. I. Plimak, Opt. Commun. \textbf{190}, 261, (2001). 
\bibitem{CDD}S. Chaturvedi, K. Dechoum, and P. D. Drummond, Phys. Rev. A \textbf{65}, 033805; P. D. Drummond, K. Dechoum and S. Chaturvedi,
Phys. Rev. A \textbf{65}, 033806 (2002). 
\bibitem{Graham}R. Graham and H. Haken, Z. Phys. \textbf{210}, 276 (1968); R. Graham \textit{ibid.} \textbf{210}, 319 (1968); \textbf{211},
469 (1968). 
\bibitem{McNeil}K. J. McNeil and C. W. Gardiner, Phys. Rev. A \textbf{28}, 1560 (1983). 
\bibitem{14}H. J. Carmichael, \textit{Statistical Methods in Quantum Optics 1}, (Springer, Berlin, 1999). 
\bibitem{+P}S. Chaturvedi, P. D. Drummond and D. F. Walls, J. Phys. A \textbf{10}, L187-L192 (1977); P. D. Drummond and C. W. Gardiner,
J. Phys. A \textbf{13}, 2353 (1980). 
\bibitem{GGD-Validity}A.~Gilchrist, C.~W.~Gardiner, and P.~D.~Drummond, Phys.~Rev.~A \textbf{55}, 3014 (1997).
\bibitem{Arnold}L. Arnold, \textit{Stochastic Differential Equations: Theory and Applications}, (John Wiley and Sons, New York, 1974); C.
W. Gardiner, \textit{Handbook of Stochastic Methods} (Springer, Berlin, 1983).
\bibitem{Wavefunction}H. J. Carmichael, \textit{An Open Systems Approach to Quantum Optics} (Springer-Verlag, New York, 1993).
\bibitem{GaugeP}P. Deuar and P. D. Drummond, Phys. Rev. A \textbf{66}, 033812 (2002).
\bibitem{Diagram}C. J. Mertens, T. A. B. Kennedy and S. Swain, Phys.~Rev.~A \textbf{48}, 2374 (1993), L. I. Plimak and D. F. Walls, Phys.~Rev.~A
\textbf{50}, 2627 (1994), C. J. Mertens and T. A. B. Kennedy, Phys.~Rev.~A \textbf{53}, 3497 (1996).
\bibitem{Yurke}B. Yurke, Phys. Rev. A \textbf{32}, 300 (1985). 
\bibitem{Gardiner}C. W. Gardiner and M. J. Collett, Phys. Rev. A \textbf{31}, 3761 (1985); M. J. Collett and D. F. Walls, Phys. Rev. A \textbf{32},
2887 (1985). 
\bibitem{Caves}C. M. Caves and B. L. Schumaker, Phys. Rev. A \textbf{31}, 3068 (1985); B. L. Schumaker and C. M. Caves, \textit{ibid.} \textbf{31}
3093 (1985). 
\bibitem{Stochdiagram}S. Chaturvedi and P. D. Drummond: Eur. Phys. J. B\textbf{8}, 251 (1999). 
\bibitem{Triple}P. D. Drummond and P. Kinsler: J. Eur. Opt. Soc. B\textbf{7}, 727 (1995); S. Chaturvedi and P. D. Drummond: Physical Review
A\textbf{55}, 912 (1997). 
\bibitem{Critical}P. Kinsler and P. D. Drummond: Phys. Rev. A\textbf{52}, 783 (1995) . 
\bibitem{crycont}T. C. Ralph, Phys. Rev. A \textbf{61}, 010303 (1999); Phys. Rev. A \textbf{62} 062306 (2000); M. Hillery, Phys. Rev. A \textbf{61},
022309 (1999); M. D. Reid, Phys. Rev. A\textbf{62}, 062308 (2000); N. J. Cerf, M. Levy and G. Van Assche, Phys. Rev. A \textbf{63}
052311 (2001); S. F. Pereira, Z. Y. Ou and H. J. Kimble, Phys. Rev. A\textbf{62}, 042311 (2000); P. Navez, E. Brambilla,
A. Gatti and L. A. Lugiato, Phys. Rev. A\textbf{65}, 013813 (2002); C. Silberhorn, T. C. Ralph, N. Lutkenhaus, G. Leuchs,
Phys. Rev. Lett. \textbf{89}, 167901 (2002); C. Silberhorn, N. Korolkova, G. Leuchs, Phys. Rev. Lett. \textbf{88}, 167902
(2002); F. Grosshans, P. Grangier, Phys. Rev. Lett. \textbf{88,} 057902 (2002). \end{thebibliography}
\end{document}